\documentclass[aps,prb,reprint,letterpaper,amsmath,amssymb, superscriptaddress]{revtex4-2}
\usepackage{soul}
\usepackage{amsmath}
\usepackage{amssymb}
\usepackage{bm}
\usepackage{epsfig}
\usepackage{graphicx}
\usepackage{color}
\usepackage[colorlinks]{hyperref}
\usepackage[T2A]{fontenc}
\usepackage[cp1251]{inputenc}
\usepackage[normalem]{ulem}
\usepackage{xcolor}
\usepackage{color}

\usepackage{empheq}
\definecolor{myblue}{rgb}{.8, .8, 1} 
\definecolor{purple}{rgb}{1,0,0.5}

\usepackage[bbgreekl]{mathbbol}

\usepackage{graphicx}
\usepackage{epstopdf}
\begin{document}
\title{Millisecond-long electron spin lifetime in CsPbI$_3$ perovskite nanocrystals revealed by optically detected magnetic resonance}
\author{Vasilii~V.~Belykh}
\email[]{vasilii.belykh@tu-dortmund.de}
\affiliation{\textit{Experimentelle Physik 2, Technische Universit\"at Dortmund, 44227 Dortmund, Germany}}
\author{Mikhail~M.~Glazov}
\affiliation{Ioffe Institute, Russian Academy of Sciences, 194021 St. Petersburg, Russia}
\author{Sergey~R.~Meliakov}
\affiliation{\textit{Experimentelle Physik 2, Technische Universit\"at Dortmund, 44227 Dortmund, Germany}}
\affiliation{\textit{P.N. Lebedev Physical Institute of the Russian Academy of Sciences, 119991 Moscow, Russia}}
\author{Dmitri~R.~Yakovlev}
\affiliation{\textit{Experimentelle Physik 2, Technische Universit\"at Dortmund, 44227 Dortmund, Germany}}
\affiliation{Ioffe Institute, Russian Academy of Sciences, 194021 St. Petersburg, Russia}
\affiliation{\textit{P.N. Lebedev Physical Institute of the Russian Academy of Sciences, 119991 Moscow, Russia}}
\author{Evgeniya~V.~Kulebyakina} 
\affiliation{\textit{P.N. Lebedev Physical Institute of the Russian Academy of Sciences, 119991 Moscow, Russia}}
\author{Mikhail~L.~Skorikov} 
\affiliation{\textit{P.N. Lebedev Physical Institute of the Russian Academy of Sciences, 119991 Moscow, Russia}}
\author{Mikhail~V.~Kochiev} 
\affiliation{\textit{P.N. Lebedev Physical Institute of the Russian Academy of Sciences, 119991 Moscow, Russia}}
\author{Maria~S.~Kuznetsova}
\affiliation{\textit{Spin Optics Laboratory, St. Petersburg State University, 198504 St. Petersburg, Russia}}
\author{Elena~V.~Kolobkova}
\affiliation{\textit{ITMO University, 199034 St. Petersburg, Russia}}
\affiliation{\textit{St. Petersburg State Institute of Technology, 190013 St. Petersburg, Russia}}
\author{Manfred Bayer}
\affiliation{\textit{Experimentelle Physik 2, Technische Universit\"at Dortmund, 44227 Dortmund, Germany}}
\affiliation{Research Center FEMS, Technische Universit\"at Dortmund, 44227 Dortmund, Germany}

\date{\today}
\begin{abstract}
Perovskite nanocrystals are a convenient model system for optical spin orientation and manipulation. However, its real potential might be underestimated due to the incomplete knowledge on spin relaxation times, which are obscured by the limited sensitivity of measurement techniques as well as by the insufficient understanding of the spin relaxation mechanisms in perovskites. In this work, we study the spin relaxation of charge carriers in perovskite nanocrystals both experimentally and theoretically. We address the electron and hole spins in CsPbI$_3$ nanocrystals embedded in a glass matrix by the resonant spin inertia technique based on optically detected magnetic resonance. It allows us to determine the longitudinal spin relaxation time $T_1$ separately for electrons and holes, the $g$~factors, and the effective Overhauser field of the nuclear spin bath. At a temperature of 1.6~K, the $T_1$ time for electrons can be as long as 0.9~ms. We reveal the effect of the time-varying nuclear field fluctuations, which enhances the electron spin relaxation at low magnetic fields, and measure a rather long nuclear spin correlation time of about 56~$\mu$s. We develop a model of the spin relaxation in nanocrystals based on a two-LO-phonon Raman process, which explains the observed temperature dependence of the time $T_1$.  
\end{abstract}
\maketitle

%\cD{As for title.   It is OK if we plan to put technique on the first place. But also in this case do we want to call it by common "ODMR technique"? Or find more specific name which reflects our development?} 
%\cD{Alternatively, we may make focus on result of millisecond times. The we should fut it on first position: (1) "Millisecond electron spin relaxation time in CsPbI$_3$ perovskite nanocrystals revealed by optically detected magnetic resonance", (2) "Millisecond electron spin relaxation time in CsPbI$_3$ perovskite nanocrystals", (3) "Millisecond long electron spin relaxation in CsPbI$_3$ perovskite nanocrystals in glass". }

%\cD{It should also differ from our previous paper in NL2022 "Submillisecond Spin Relaxation in CsPb(Cl,Br)3 Perovskite Nanocrystals in a Glass Matrix". While seems that it is fulfilled, as one order is added, electrons are specified, material is  different...}
%\commV{The goal of the paper is not to introduce a technique, resonant spin inertia, as it was introduced back in 2022. As for me, in the title it is important to have ``optical'', and ``millisecond''. ODMR is recognised by people and also not so many reports on ODMR in perovskites, so let's make indeed ``Millisecond electron spin relaxation time in CsPbI$_3$ perovskite nanocrystals revealed by optically detected magnetic resonance''.}

\section{Introduction}
Lead halide perovskite semiconductors have recently attracted great attention as a promising material for solar cells \cite{Liu2013,Green2014,han2025perovskite}. This has stimulated intense investigations of perovskites beyond photovoltaics that revealed potential for their application as X-ray detectors \cite{Stoumpos2013,Yakunin2015}, light-emitting diodes \cite{Cao2018,Liu2021,Kovalenko2017}, quantum emitters \cite{Utzat2019}, etc. Lead halide perovskites are optically active (direct band gap) semiconductors with transition energies lying in the visible or near infrared range. They can be synthesized in the form of bulk single crystals with a well defined optical transition energy or nanocrystals (NCs), where the transition energy (and emission wavelength) can be tailored by varying the NC size and composition \cite{Protesescu2015}. Perovskite NCs embedded in glass matrix \cite{Kolobkova2021} as studied in this work demonstrate the robustness with respect to the environment.

Perovskites demonstrate interesting spin properties, which can be studied by the optical techniques developed for traditional semiconductors \cite{Nestoklon2018,Odenthal2017,Belykh2019, Crane2020, Grigoryev2021,Kirstein2022Uni,Nestoklon2023,Kirstein2025RSA,Meliakov2025IE,zhu2024coherent,tamarat2019ground}. Unlike in prototypical GaAs, both the electrons and the holes at the bottom of the conduction band and the top of the valence band, respectively, carry spin $1/2$ \cite{Yu2016}; perovskites feature a stronger hyperfine interaction with the nuclear spins for the holes compared to that for the electrons \cite{Kirstein2022Adv,kotur2026dynamic}. They show nanosecond spin coherence times at room temperature \cite{lin2023room,Meliakov2023}, also because of the lack of the Dyakonov-Perel relaxation mechanism, as confirmed by strong optical orientation under nonresonant excitation \cite{Kopteva2024,Kopteva2025}. 

The possibility of practical applications of a spin system depends on the values of the spin lifetimes $T_1$ and $T_2$.
The spin coherence time $T_2$, which characterizes the unperturbed spin precession about a magnetic field, determines the applicability of a spin state as a qubit. On the other hand, the relaxation time $T_1$ of a spin along the magnetic field sets the ultimate limit for $T_2$, when it is extended by decoupling protocols. In perovskites, $T_1$ and $T_2$ are only little explored mainly due to limitations of the applied techniques. The spin coherence time $T_2$ is obscured by the spread of $g$ factors and by nuclear field fluctuations in pump--probe experiments. As a result, these experiments reveal the inhomogeneous spin dephasing time $T_2^*$ of a few ns rather than the spin coherence time $T_2$ \cite{Crane2020,Grigoryev2021}. On the other hand, longitudinal spin dynamics, measured by polarization-resolved photoluminescence, can be limited by the carrier lifetime, rather than by $T_1$ \cite{Kopteva2024,Kopteva2025}. 
The extended pump--probe Faraday/Kerr experiments, which can access also long-living resident carriers, show that the longitudinal spin dynamics is nonexponential and only spin lifetime corresponding to its fast component of about~100 ns was determined so far \cite{Kirstein2022Adv,Kudlacik2024}. Measurements over longer time scales are complicated by the weak signal resulting from the low repetition rates of the laser pulses. 
Furthermore, standard measurements of the longitudinal spin relaxation (photoluminescence dynamics with polarization control and pump-probe experiments) cannot answer %\cD{clarify or help to clarify} 
whether the obtained relaxation time should be attributed to electrons or to holes. Recently, a novel technique that exploits optically detected magnetic resonance (ODMR) with detection of the Faraday rotation combined with the spin inertia effect was developed and demonstrated for rare earth Ce$^{3+}$ ions in YAG matrix \cite{Belykh2022PRB}. This technique, referred to as resonant spin inertia, makes it possible to measure the time $T_1$ for carriers with a distinct value of the $g$ factor and, in particular, attribute $T_1$ to electrons or holes. At the same time, it was applied to CsPb(Cl,Br)$_3$ NCs at low magnetic fields, where a submillisecond-long time $T_1$ was measured and attributed to the spatially indirect exciton \cite{Belykh2022NL}. On the other hand, there is a limited knowledge on the long-time spin dynamics for non-interacting electrons and holes in perovskite NCs beyond the work \cite{barak2022uncovering}, where submillisecond spin lifetimes were reported for electrons and holes in CsPb(Cl,Br)$_3$ NCs. For independent charge carriers one expects even longer spin relaxation times because for excitons the electron-hole exchange interaction can control its spin dynamics. The detailed understanding of the spin relaxation mechanisms in perovskite NCs, in particularly those in embedded in a glass matrix and demonstrating outstanding stability, may have potential for quantum informatics applications of this spin system.

In this study, we apply this ODMR-based resonant spin inertia technique to CsPbI$_3$ NCs, clearly observe electron and hole resonances, and measure the corresponding longitudinal spin relaxation times $T_1$. We show that the electron $T_1$ increases with magnetic field due to the suppression of the effect of fluctuating nuclear fields on the electron spin relaxation. This dependence makes it possible to estimate the characteristic time of the nuclear field variations, which appears to be much longer than that reported for, e.g., GaAs quantum dots \cite{Bechtold2015}. For relatively high magnetic fields, low laser powers, and a temperature of 1.6~K, we find the electron $T_1$ to be as long as 0.9~ms, which, to the best of our knowledge, is the longest value reported for perovskite NCs. The temperature dependence of $T_1$ shows an activation-like behavior with a characteristic energy corresponding to a low-frequency LO phonon energy in the CsPbI$_3$ perovskite. We develop a theory of spin relaxation in perovskite NCs based on a two-phonon Raman process, which accounts for the experimental findings. As an additional result of this work, we also observe an increase in the electron spin resonance linewidth with increase in the transition energy, which is explained by the increasing nuclear fluctuations for smaller NCs.

\section{Experimental details}

The samples under study are CsPbI$_3$ NCs embedded in a fluorophosphate Ba(PO$_3$)$_2$-AlF$_3$ glass matrix. 
They were synthesized by rapid cooling of a glass melt enriched with the components needed for perovskite crystallization as described in Refs. \cite{Kirstein2023,Kolobkova2021,Meliakov2025IE,Meliakov2024IT}. Two samples, in the following referred to as \#1 and \#2 (technological codes EK202 and EK8-M), are studied. The average NCs size for the sample \#1 is smaller than that for the sample \#2, while in general for the two samples NC sizes range from 7 to 14~nm. The two samples show qualitatively a similar behavior of their spin properties, all results except the spectral dependences (Fig.~\ref{fig:EDep}) are shown for the sample \#1. The sample is placed into a Helium-bath cryostat with a superconducting magnet allowing us to apply a magnetic field $\textbf{B}$ in the Faraday geometry (along the laser beam). All experiments are performed at a temperature of $T = 1.6$~K unless otherwise stated. 

We measure the optically detected magnetic resonance stimulated by optical pumping and evaluate the longitudinal spin relaxation time $T_1$ with the technique described in Refs.~\cite{Belykh2022PRB,Belykh2022NL}. The sample is excited by the radiation of a Coherent Chameleon Discovery laser system emitting 100~fs pulses (7~nm spectral width) with a repetition rate of 80~MHz, tunable across a wide spectral range. Note that the fact that the laser is pulsed (rather than continuous wave) is not essential for this technique as long as the separation between pulses is much smaller than the time $T_1$. The laser wavelength is tuned into resonance with the fundamental optical transition of the studied NCs. The polarization of the laser beam passing the sample is set to be elliptical, so that the circularly polarized component of the beam excites carrier spin polarization in the sample. This polarization is detected by measuring Faraday rotation experienced by the linearly polarized component. The Faraday rotation was measured using a conventional scheme: the transmitted laser beam passed a half-wave plate followed by a Wollaston prism, splitting it into two beams with orthogonal linear polarizations whose intensities are then measured by a balanced photodetector. The difference signal from the photodetector is proportional to the Faraday rotation angle and, ultimately, to the spin polarization in the sample. A coil with a diameter of about 1~mm is placed near the sample surface, so that the laser beam passes through both the sample and the coil. The coil is used to apply a radiofrequency (rf) magnetic field with frequency up to 4~GHz to the sample. 

Thus, the spin polarization is created and detected using the optical beam, while the rf field applied in resonance with the Larmor spin precession frequency destroys the spin polarization \cite{Belykh2022PRB}. The rf power is modulated at the frequency $f_\text{mod}$ ranging from 0.1 to 100~kHz and the corresponding modulation of the spin polarization is detected with the balanced photodetector and a lock-in amplifier. By scanning the magnetic field with the rf frequency $f_\text{rf}$ fixed, it is possible to detect the magnetic resonances for electrons and holes in the studied NCs. 
It is also possible to evaluate the longitudinal spin relaxation time $T_1$. Indeed, the signal is independent of the rf modulation frequency $f_\text{mod}$ for low frequencies, where the modulation period is longer than the time $T_1$. At higher frequencies, $f_\text{mod}\gtrsim 1/T_1$, the signal begins to decrease with $f_\text{mod}$, and so the modulation-frequency dependence of the signal makes it possible to evaluate $T_1$. This is similar to the all-optical spin inertia method \cite{Heisterkamp2015}, but in contrast to it, in our case $T_1$ can be measured selectively for a given spin resonance (e.g., the electron or the hole) by setting the value of the rf frequency or the magnetic field; correspondingly, this method is referred to as resonant spin inertia \cite{Belykh2022PRB}.  

\section{ODMR spectra of perovskite nanocrystals}
\begin{figure*}
\includegraphics[width=1.9\columnwidth]{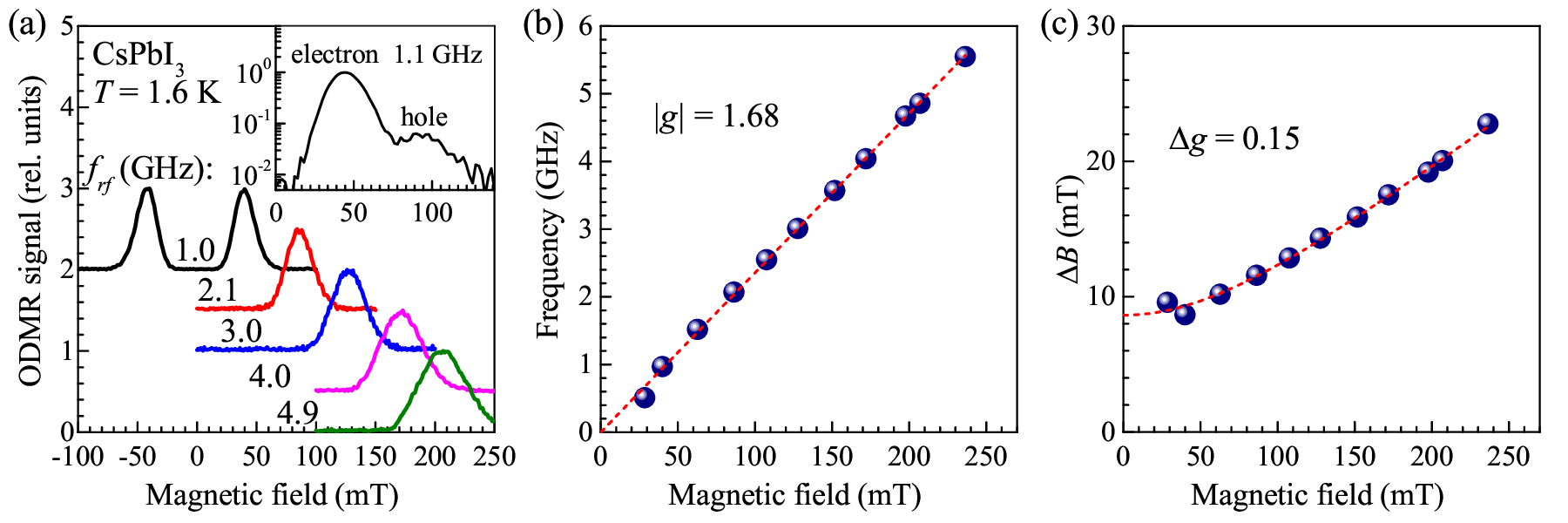}
\caption{Optically detected magnetic resonance in CsPbI$_3$ NCs. (a) ODMR spectra for different rf field frequencies. The modulation frequency is 0.6~kHz, the laser photon energy is 1.824~eV, the laser power is 0.5~mW. The inset shows the ODMR spectrum for $f_\text{rf} = 1.1$~GHz at an increased modulation frequency of 3~kHz and a laser power of 2~mW manifesting two resonances. (b) Magnetic field dependence of the magnetic resonance frequency  allowing one to determine the electron $g$ factor of 1.68 from the linear fit shown by the red dashed line. (c) Magnetic field dependence  of the width of the electron ODMR resonance (defined as the normal distribution standard deviation $\sigma$). The red dashed line shows a fit with Eq.~\eqref{eq:dB}. The temperature is 1.6~K.}
\label{fig:BScan}
\end{figure*}

Figure~\ref{fig:BScan}(a) shows the dependence of the Faraday rotation signal on the magnetic field (ODMR spectrum) for different rf field frequencies. The spectrum at $f_\text{rf} = 1.0$~GHz is measured both in the positive and negative range of magnetic fields and is rather symmetric, which indicates the absence of %internal magnetic fields in the system, in particular the lack of 
a notable dynamic nuclear polarization. The spectra show a clear resonant behavior of the Larmor spin precession frequency and the rf field frequency $f_\text{rf}$, described by the equation 
\begin{equation}
2\pi\hbar f_\text{rf} = |g| \mu_\text{B} B,
\label{eq:Larmor}
\end{equation}
where $g$ is the $g$ factor and $\mu_\text{B}$ is the Bohr magneton.
As the rf field frequency is increased, the resonance peak shifts towards higher fields and broadens. Figure~\ref{fig:BScan}(b) shows the resonance frequency as a function of the magnetic-field position of the peak. The dependence is linear with the slope corresponding to a $g$ factor of $1.68\pm 0.02$. The value of this $g$ factor at the energy of 1.824~eV indicates that it belongs to electrons and its sign is positive, according to %universal dependence of the $g$ factor on the band gap in perovskites 
theoretical predictions \cite{Nestoklon2023}. 

The ODMR peak can be well fitted by a Gaussian distribution with the standard deviation $\sigma$ giving the width of the ODMR peak $\Delta B$. The width increases with growing magnetic field [Fig.~\ref{fig:BScan}(c)]. This increase is related to the spread of $g$ factors $\Delta g$ \cite{belykh2016large} and can be described by the equation
\begin{equation}
\Delta B = \left[\Delta_\text{N}^2/2 + \left(\frac{\Delta g}{g} B \right)^2\right]^{1/2}.
\label{eq:dB}
\end{equation}
This equation assumes that the finite width of the ODMR peak is contributed by the spread of Larmor frequencies resulting from the Gaussian distribution of $g$ factors with standard deviation $\Delta g$ and by the Gaussian distribution of the effective Overhauser field $\mathbf{B}_\text{N}$ of the nuclear spin fluctuations $\pi^{-3/2}\Delta_\text{N}^{-3}\exp(-\mathbf{B}_\text{N}^2/\Delta_\text{N}^2)$ with $\Delta_\text{N}$ being the spread of the nuclear field fluctuations \cite{merkulov2002electron,glazovElectronNuclearSpin2018}. The $\Delta g$ contribution can be evaluated from Eq.~\eqref{eq:Larmor}. The convolution of these two Gaussian distributions results in a Gaussian distribution with the standard deviation given by Eq.~\eqref{eq:dB}. Fitting the dependence of $\Delta B$ on $B$ [Fig.~\ref{fig:BScan}(c)] with Eq.~\eqref{eq:dB}, we determine $\Delta g = 0.15$ and $\Delta_\text{N} \approx 12$~mT. These values are in good agreement with those obtained from pump--probe Faraday rotation measurements on a similar sample by analyzing the electron spin precession in zero field and the precession decay in high fields \cite{Meliakov2026Nucl}.

Note that apart from the strong electron resonance, the ODMR spectra reveal a much weaker peak at higher fields [inset in Fig.~\ref{fig:BScan}(a)], corresponding to $|g| \approx 0.8$, related to holes \cite{Nestoklon2023,Meliakov2025IE,Meliakov2024IT}. The measured electron and hole $g$ factors are in good agreement with those measured by the pump-probe Faraday rotation technique for a similar sample \cite{Meliakov2025IE}. The width of the hole peak $\Delta B \approx 20$~mT is larger than that of the electron peak, indicating stronger hyperfine interaction, which is characteristic of holes in perovskites \cite{Kirstein2022Adv}. Interestingly, the hole peak is observed only in a finite range of rf field frequencies, which is characteristic of the carriers within an exciton \cite{Belykh2022NL}.

\begin{figure}
\includegraphics[width=0.9\columnwidth]{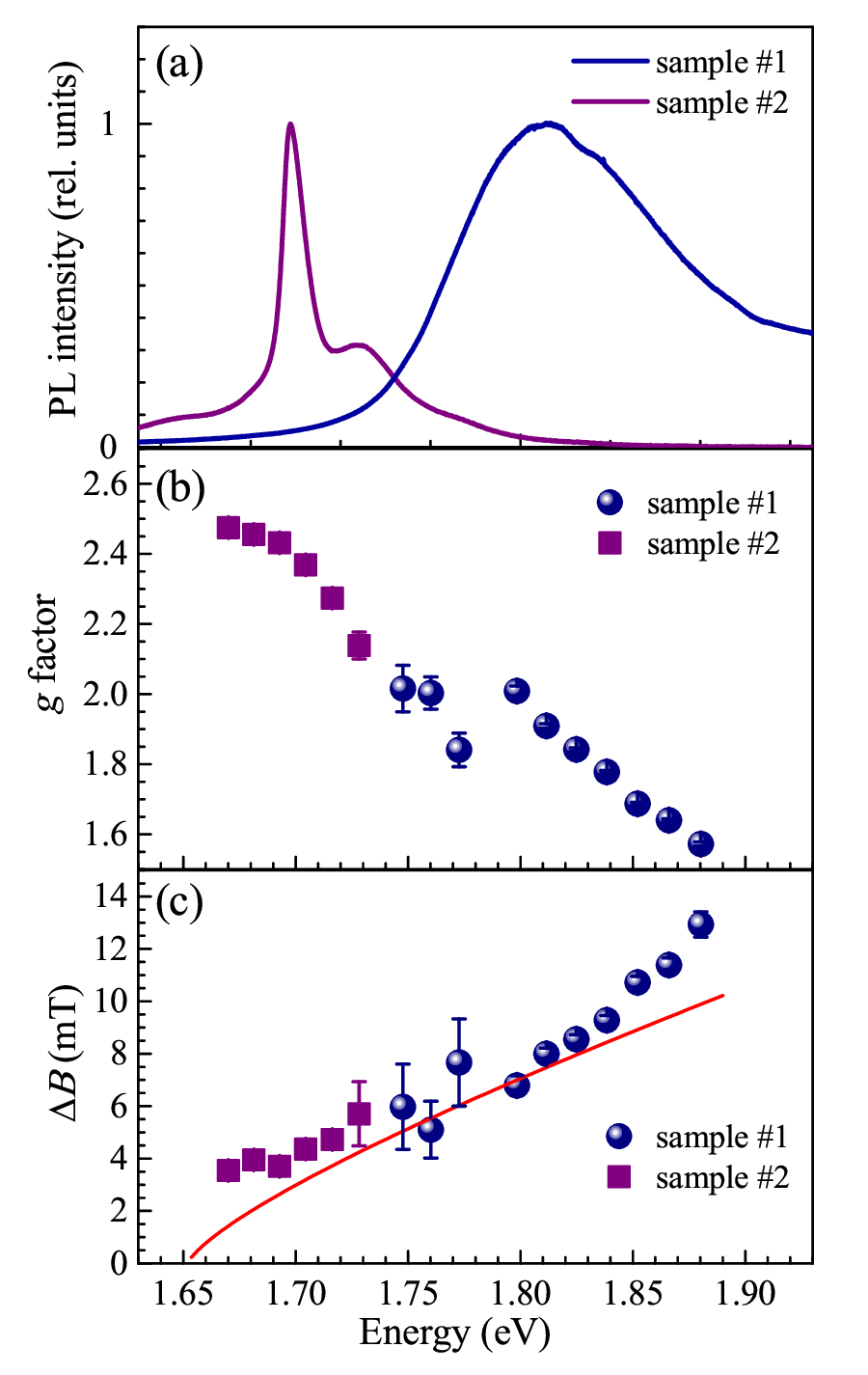}
\caption{(a) Photoluminescence spectra of the two CsPbI$_3$ perovskite samples with different average NC sizes measured at $T = 6$~K. Dependence of electron $g$ factor on optical transition energy (b) and of electron ODMR peak width (c) measured at $f_\text{rf}$ = 1.13 GHz (magnetic field of about 45~mT). The red solid line shows the calculated nuclear field according to Ref.~\cite{Meliakov2026Nucl}, see text. The temperature in (b) and (c) is 1.6~K.}
\label{fig:EDep}
\end{figure}

Next, we present the dependence of the electron $g$ factor and effective nuclear field on the NC size parametrized by the optical transition energy. The sample contains an inhomogeneous ensemble of NCs with different sizes and, thus, different quantum-confinement energies resulting in a spread of their emission energies. We can address NCs of different sizes within the same sample by changing the laser photon energy. Higher energy corresponds to NCs of smaller size. Here we take an advantage of the fact that the Faraday rotation signal sharply increases when laser energy is resonant with optical transition energy (band gap if we neglect excitonic effect) and vanishes when the laser energy is detuned by several meV, as it was shown in Refs.~\cite{Belykh2019,Kirstein2022Adv}.

We study two samples with different average sizes of NCs to address a broader range of energies. The optical transition energy dependence of the $g$ factor is shown in Fig.~\ref{fig:EDep}(b), while for reference we present the photoluminescence (PL) spectra of the samples in Fig.~\ref{fig:EDep}(a). The electron $g$ factor decreases with growing energy, which corresponds to the general trend of the universal $g$ factor dependence on the bandgap \cite{Kirstein2022Uni}, confirming the electron origin of the observed resonance. A similar energy dependence of the $g$ factor measured by a pump--probe technique for CsPbI$_3$ perovskite NCs is discussed in detail in Refs.~\cite{Nestoklon2023,Meliakov2025IE}. 

Figure~\ref{fig:EDep}(c) shows the corresponding energy dependence of the ODMR peak width $\Delta B$ (standard deviation) at a rather small field of about 45~mT. This width mostly originates from the effective nuclear field, $\Delta B \approx  \Delta_\text{N}/\sqrt{2}$, as one can see from Fig.~\ref{fig:BScan}(c) and Eq.~\eqref{eq:dB}. It follows from this figure that the nuclear field increases with energy, i.e., with a decreasing NC size $a$. This behavior is expected, as the number of nuclear spins $N \propto a^3$  in a NC. The net spin in a NC averaged over all nuclei having random spin orientation is proportional to $N^{1/2} \propto a^{3/2}$. On the other hand, the electron wavefunction spreads over the entire NC volume, and the contribution of an individual nuclear spin to the hyperfine field sensed by the electron is proportional to $a^{-3}$.  As a result, the total hyperfine field is proportional to $a^{3/2}a^{-3} \propto a^{-3/2}$. The proportionality coefficient depends on the hyperfine constants and the contributions of the different nuclear species. It was calculated in Ref.~\cite{Meliakov2026Nucl} for the same system. Using these calculations with the hyperfine constants of Pb and I of 50 and 190~$\mu$eV, respectively, we get $\Delta B \approx \Delta_\text{N}/\sqrt{2} = (a_0/a)^{3/2} \times (0.47\text{ T})$, where $a_0 = 0.624$~nm is the lattice constant. To relate the transition energy to the NC size, we use Eq.~(S10) from the Supporting Information of Ref.~\cite{Harkort2025}. The resulting calculated curve %\cD{dependence?} 
is shown by the red solid line in Fig.~\ref{eq:dB}(c). It shows reasonable agreement with the experimental dependence. The deviation at small energies is related to the limited accuracy of the NC size estimation when the NC transition energy is close to that of bulk material.
%\CommMS{One one hand, the discrepancy at high energies is no smaller that at low energies. On the other hand, the energy range for sample 2 below 1.7 eV most probably corresponds to defect/impurity states, and their spin properties may be quite different from the bulk of NCs.}

\section{Longitudinal spin relaxation time, $T_1$: resonant spin inertia experiments}
\begin{figure*}
\includegraphics[width=1.9\columnwidth]{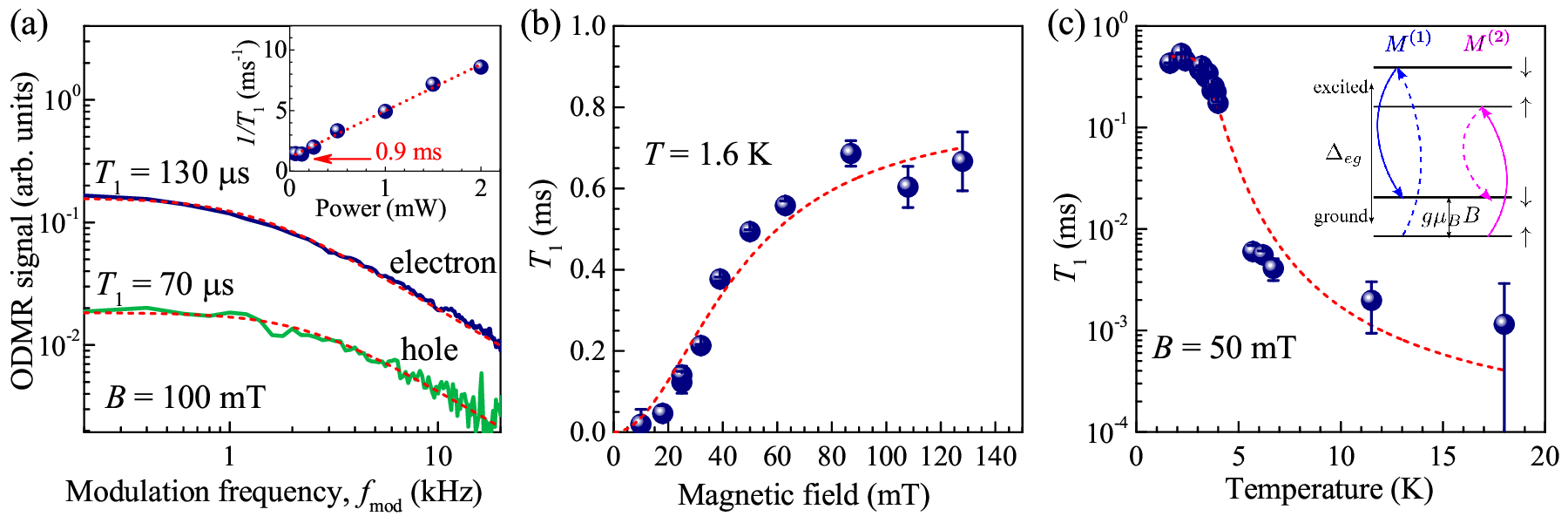}
\caption{(a) Dependence of the ODMR signal on the modulation frequency for the electron and hole resonances for a magnetic field of 100~mT (the corresponding rf frequencies are 2.2 and 0.6~GHz) and an increased laser power of 2~mW. The curves demonstrate the spin inertia effect making it possible to determine the times $T_1$ for the electrons and the holes separately. The dashed lines show fits with Eq.~\eqref{eq:SI}. The inset shows the dependence of $1 / T_1$ on the laser power for the electron with a linear fit shown by the dashed line. (b) Magnetic field dependence of the electron $T_1$ measured with a laser power of 0.125~mW. The dashed line shows a fit with Eq.~\eqref{eq:T1}. (c) Temperature dependence of the electron $T_1$ corresponding to a laser power of 0.125~mW. The dashed line shows a fit with an activation dependence~\eqref{eq:act}. The inset shows a sketch of virtual transitions to the NC exited state, accompanied by the emission and absorption of LO phonons resulting in a spin flip. The laser photon energy is 1.824~eV. The temperature in panels (a) and (b) is 1.6~K.}
\label{fig:T1}
\end{figure*}

Here we turn to the measurement of the longitudinal spin relaxation time $T_1$. In this experiment, we fix the magnetic field, so that the rf field is resonant with the electron (or hole) Larmor frequency and measure the ODMR signal as a function of the rf field modulation frequency $f_\text{mod}$ (spin inertia curve). Figure~\ref{fig:T1}(a) shows the spin inertia curves for the electron and hole resonances. The signal weakly depends on $f_\text{mod}$  at low frequencies, when $f_\text{mod} \ll 1/T_1$. As the frequency is increased above $1/T_1$, the accumulated spin, i.e. the signal, becomes limited by the modulation period $1/f_\text{mod}$ rather than $T_1$. This is described by the formula \cite{Belykh2022PRB}
\begin{equation}
S = \frac{S_0}{\sqrt{1 + (2 \pi T_1 f_\text{mod})^2}},
\label{eq:SI}
\end{equation}
where $S_0$ is the amplitude of the spin signal $S$ under infinitely slow modulation. Fitting the experimental curves with this equation [red dashed lines in Fig.~\ref{fig:T1}(a)] gives $T_1 = 130$~$\mu$s for the electrons and 70~$\mu$s for the holes. Note that these measurements are performed for an increased excitation power of 2~mW to resolve the hole resonance. At the same time, a decrease in the laser power leads to an increase in $T_1$ as depicted in the inset of Fig.~\ref{fig:T1}(a). The experimental dependence suggests that the rate $1/T_1$ increases linearly with power $P$, and extrapolation of the dependence to zero power gives $T_1 = 0.9$~ms. The decrease of $T_1$ with increasing laser power is characteristic for many other systems studied optically \cite{Heisterkamp2015,Belykh2022PRB,Belykh2022NL}. It may be related to the re-excitation  of already spin-oriented NCs.
%The remarkably long $T_1$ times of electrons and holes evidence that the measured signals originate from singly-charged NCs, either by an electron or by a hole. This is in line with our previous studies of CsPbI$_3$ and CsPbBr$_3$ NCs in glass investigated by means of time-resolved Faraday ellipticity [Nanoscale 17, 6522 (2025); Nanoscale 16, 21496 (2024)]. The resident carriers in the NCs can arise from long-living photocharging, where either the electron or the hole from a photogenerated electron-hole pair escapes from the NC and or is captured by the surface state. As a result, some NCs in the ensemble are charged with electrons, some NCs with holes, while the rest remain neutral.
The remarkably long $T_1$ times of electrons and holes evidence that the measured signals originate from long-living (resident) charge carriers in NCs. These carriers can arise from photogenerated electron-hole pairs, when one of the carriers is captured by the surface trap, which prevents subnanosecond recombination \cite{kulebyakina2024temperature}, while the other carrier can be oriented and detected optically similarly to optical orientation of singly charged quantum dots described in Ref.~\cite{Greilich2006}. We do not exclude formation of the spatially indirect long-living excitons in the case of shallow traps.

The dependence of the electron time $T_1$ on the magnetic field is shown in Fig.~\ref{fig:T1}(b). One can see that $T_1$ increases strongly with $B$ within 60~mT and then saturates. This is the typical behaviour expected when $T_1$ is determined by fluctuating internal (nuclear) magnetic fields, whose impact is suppressed by the external magnetic field. This scenario was considered in Ref.~\cite{Smirnov2018}, where the following analytical formula for $T_1(B)$ was proposed:
\begin{equation}
T_1 = \frac{\tau_\text{s}}{1 + (\Delta_\text{N} / B)^2 \tau_\text{s} / \tau_\text{c}}.
\label{eq:T1}
\end{equation}
Here, $\tau_\text{s}$ is the spin relaxation time limited by mechanisms other than the nuclear-field fluctuations, $\tau_\text{c}$ is the correlation time of the nuclear fluctuations characterizing evolution of the effective nuclear field within NC. This equation is valid when $B$ exceeds the spread of nuclear fields $\Delta_\text{N}$ and the corresponding Larmor spin precession frequency exceeds $1/\tau_\text{c}$. Both conditions are satisfied in our experiments. For $B \gg \Delta_\text{N}$, Eq.~\eqref{eq:T1} yields $T_1 \approx \tau_\text{s} \approx 0.8$~ms. We can estimate $\Delta_\text{N} \approx 12$~mT from the width of the ODMR resonance for $B \rightarrow 0$ [Fig.~\ref{fig:BScan}(c) and Eq.~\eqref{eq:dB}]. Fitting the dependence $T_1 (B)$ with Eq.~\eqref{eq:T1} [red dashed lines in Fig.~\ref{fig:T1}(b)], we then obtain $\tau_\text{c} = 56 \pm 9$~$\mu$s, $\tau_\text{s} = 790\pm 60$~$\mu$s. The time $\tau_\text{c}$, characterizing the random variation of the effective nuclear field in NCs, is longer than that reported for (In,Ga)As QDs ($\sim$~10 $\mu$s) \cite{Bechtold2015}. We note that in CsPbI$_3$ NCs the hyperfine interaction for electrons is dominated by the iodine nuclei \cite{Meliakov2026Nucl}, which have spin $5/2$, enabling their quadrupole interaction. The latter seems to be weaker than, e.g., in GaAs-based systems, as suggested by the long $\tau_\text{c}$. We also note that the difference in the hyperfine interaction parameters for electrons and holes, resulting in the different $\Delta_\text{N}$, can explain their difference in $T_1$ presented in Fig.~\ref{fig:T1}(a) on the basis of Eq.~\eqref{eq:T1}.

We note that in the studied CsPbI$_3$ NCs, the $T_1$ time becomes quite short in weak magnetic fields. For $B = 10$~mT, it is about $20$~$\mu$s, while at $B = 0$~mT the signal decreases to the noise level, indicating even shorter $T_1$ (the signal amplitude is proportional to $T_1^2$ \cite{Belykh2022PRB}). 
Somewhat different behavior was observed in Ref.~\cite{Belykh2022NL} for CsPb(Cl,Br)$_3$ NCs, where the spin resonance was attributed to a spatially indirect exciton with a differential electron-hole $g$ factor of electron and hole. In the latter case, $T_1$ has a finite value of about $30$~$\mu$s already at $B=0$ which can be explained by a certain stabilization of the exciton spin against nuclear fluctuations due to electron-hole exchange interaction. We note that the magnitude of the exchange interaction and, consequently, the robustness of $T_1$ against nuclear fluctuations at $B=0$, depends on the type of surface traps that localize the carriers and determine the overlap of their wave functions. The nature of the traps may depend on the NCs composition and size, which differ between the present work and Ref.~\cite{Belykh2022NL}. At magnetic fields exceeding the effective exchange field, the carriers bound in an exciton exhibit behaviour similar to that of independent carriers.

The dependence of the electron time $T_1$ on temperature $T$ is shown in Fig.~\ref{fig:T1}(c). This dependence is weak below 3~K, and then $T_1$ starts to decrease rapidly. This decrease can be described by an activation dependence, which might take place if relaxation is caused by activation to some energy level at an energy $E_\text{a}$ above the considered state: 
\begin{equation}
\frac{1}{T_1} = \frac{1}{T_{1,0}} + \gamma_\text{a} \exp\left(-\frac{E_\text{a}}{k_\text{B}T}\right). 
\label{eq:act}
\end{equation}
Here $T_{1,0}$ is the relaxation time in the low-temperature limit and $\gamma_\text{a}$ is the activation rate, $k_\text{B}$ is the Boltzmann constant. The red dashed line in Fig.~\ref{fig:T1}(c)] shows a fit of the experimental data by this equation with parameters $E_\text{a} = 2.8\pm 0.2$~meV and $\gamma_\text{a} = 15\pm 8$~$\mu$s$^{-1}$. We note that this value of $E_\text{a}$ is close to an optical-phonon energy of 3.2 meV \cite{Harkort2025,Trifonov2026}, which suggests that the spin relaxation process involves optical phonons. This will be confirmed by the theoretical considerations below. We note that strictly speaking we measure spin lifetime, which can be contributed by the carrier lifetime and by the time $T_1$. However strong dependence of measured time on magnetic field and temperature indicates that in our case spin lifetime is dominated by $T_1$, and we probe the NCs subensemble containing resident carriers. Nevertheless, we do not exclude that at lowest temperatures and high magnetic fields spin lifetime may have notable contribution from carrier lifetime.

\section{Theory of two-optical phonon induced electron spin flip in perovskite nanocrystals}
Let us discuss the temperature dependence of the longitudinal spin relaxation time $T_1$ in more detail. We note that real activation to an excited state in NCs under study is hardly possible at low temperatures, since the corresponding energy separation is in the tens of millielectronvolts range. Thus, one has to consider virtual transitions in second-order perturbation theory involving phonons, which provide an effective Rashba-like spin--orbit coupling. For a model description, we consider a NC in the form of a cuboid with principal axes $x$, $y$, $z$ that also correspond to the crystallographic axes. Here we disregard the reduced symmetry of the perovskite host and consider it to be in the cubic $O_h$ phase. Let the magnetic field $\bm B\parallel z$ be directed along one of its principal axes. We assume the interaction of charge carriers (electrons for definiteness) with bulk optical phonons via the Fr\"ohlich (polarization) mechanism. 

To that end, we recall that a longitudinal optical (LO) phonon with wavevector $\bm q$ produces an electric potential
\begin{equation}
\label{LO:Fro}
\varphi_{\bm q} = \alpha_F \frac{\mathrm i}{q} \left(\hat{b}_{\bm q} e^{\mathrm i \bm q \bm r - \mathrm i \Omega_q t} - \hat{b}_{\bm q}^\dag e^{-\mathrm i \bm q \bm r + \mathrm i \Omega_q t}\right),
\end{equation}
where $\hat{b}_{\bm q}$ and $\hat{b}_{\bm q}^\dag$ are the phonon annihilation and creation operators, respectively, $\Omega_q$ is the phonon frequency, and $\alpha_F$ is the Fr\"ohlich interaction constant related to the macroscopic properties of the material. In the simplest model with one type of optical phonons 
\begin{equation}
\label{alpha:F}
\alpha_F = - \left(\frac{2\pi \hbar\Omega}{\varepsilon^* \mathcal V} \right)^{1/2}, \quad \frac{1}{\varepsilon^*} = \frac{1}{\varepsilon_\infty} - \frac{1}{\varepsilon_\text{s}},
\end{equation}
where $\mathcal V$ is the normalization volume, $\varepsilon_\text{s}$ and $\varepsilon_\infty$ are the static and high-frequency dielectric constants, respectively, and the dependence of $\Omega$ on $q$ is neglected. %\commV{Maybe let's move definition of $\alpha_F$ to the appendix.}
Hence, LO phonons create an electric field
\begin{equation}
\label{E:Fro}
\bm{\mathcal E}_{\bm q} = -\bm \nabla \varphi_{\bm q} = \alpha_F \frac{\bm q}{q} \left(\hat{b}_{\bm q} e^{\mathrm i \bm q \bm r - \mathrm i \Omega_q t} + \hat{b}_{\bm q}^\dag e^{-\mathrm i \bm q \bm r + \mathrm i \Omega_q t}\right).
\end{equation}
By virtue of spin--orbit interaction, this electric field induces a Rashba-like spin--orbit coupling that results in a spin--phonon interaction described by
\begin{equation}
\label{spin:phon}
\mathcal H_{sf} = \alpha_R \left\{[\hat{\bm k} \times \bm{\mathcal E}]\right\}_{\rm sym} \cdot \hat{\bm \sigma},
\end{equation}
where $\alpha_R$ is the Rashba constant related to the spin-orbit interaction,  $\hat{\bm k}= - \mathrm i \bm \nabla$ is the electron wavevector operator, and $\hat{\bm \sigma}$ is the pseudovector composed of the spin-Pauli matrices representing the electron spin $\hat{\bm s} = \hat{\bm \sigma}/2$. The curly brackets in Eq.~\eqref{spin:phon} stand for the symmetrization of the operators $\{AB\}_{\rm sym} = (AB+BA)/2$ since, generally, $\hat{\bm k}$ does not commute with the coordinate-dependent electric field. Equation~\eqref{spin:phon} represents the so-called direct spin--phonon coupling~\cite{pavlovSpinFlipInteraction1966,pavlovSpinphononInteractionElectrons1967,glazovElectronNuclearSpin2018}. 

%Let $\bm q_1$ and $\bm q_2$ be the wavevectors of the two phonons. %Symmetry-wise the spin-dependent part of the two-phonon matrix element has the following form
%\begin{equation}
%\hat{M}_{sf} = V \frac{[\bm q_2 \times \bm q_1]\hat{\bm \sigma}}{q_2q_1} ,
%\end{equation}
%where $V$ is the parameter which smoothly depends on $q_1$ and $q_2$ and non-zero at $q_1,q_2\to 0$ limit. It is determined by the microscopic process, see below.
%
The two-phonon contribution to the spin-flip rate at low temperatures $k_B T \ll \hbar\Omega_q$ is related to a process, where one phonon is absorbed and another one is emitted (Raman-like process). These processes are accompanied by virtual transitions of electrons from the ground (g) to the excited (e) state of the NC and back. A sketch of these processes is shown in the inset of Fig.~\ref{fig:T1}(c), where two contributions to the spin-flip matrix element are seen:
\begin{equation}
\label{me:2}
M_{\downarrow\uparrow} = M^{(1)}+M^{(2)}.
\end{equation} 
The process where the upwards transition is accompanied by an electron spin flip, while the subsequent downwards transition is spin-conserving is described by the term $M^{(1)}$; the opposite case is described by the term $M^{(2)}$. We note that optical phonons in NCs are confined due to the mismatch between the dielectric properties of the crystal and the matrix. The details and specific models can be found, e.g., in Refs.~\cite{englmanOpticalLatticeVibrations1968a,ruppinOpticalPhononsSmall1970,efrosElectronHolePairPhonon1993}. The electron--phonon interaction is described in Refs.~\cite{kleinSizeDependenceElectronphonon1990,efrosElectronHolePairPhonon1993}. However, for estimates, we use the long-wavelength approximation assuming that $q\ll 1/a$, where $a$ is the typical NC size. Although this approximation may not be fulfilled in experiment, it allows us to obtain a compact and reasonable analytical estimate. As a result, we can neglect the coordinate dependence of the field $\mathcal E$ in Eqs.~\eqref{E:Fro} and \eqref{spin:phon} and use the spin-independent interaction in the form
\begin{equation}
\mathcal H_{orb} = - e\bm{\mathcal E} \cdot \bm r.
\label{eq:Horb}
\end{equation}

Let us now calculate the total matrix element~\eqref{me:2} assuming that initially we have an electron in the state $g\uparrow$ and a phonon with wavevector $\bm q_1$, while in the final state we have an electron in the state $g\downarrow$ and a phonon with wavevector $\bm q_2$. Hence, the phonon $\bm q_1$ is absorbed and the phonon $\bm q_2$ is emitted. We then have
%\begin{widetext}
\begin{multline}
\label{eq:M1}
M^{(1)}(\bm q_1,\bm q_2) = -M^{(2)}(\bm q_2,\bm q_1) =\\
-\frac{M^{(em)}_{ge}(\bm q_2)M^{(abs, sf)}_{eg}(\bm q_1)}{\Delta_{eg}-\hbar\Omega_1}-\frac{M^{(abs)}_{ge}(\bm q_1)M^{(em, sf)}_{eg}(\bm q_2)}{\Delta_{eg}+\hbar\Omega_2}.
\end{multline}
%\end{widetext}
%Note that $M^{(1)}(\bm q_1, \bm q_2) = -M^{(2)}(\bm q_2,\bm q_1)$.
Here, $M^{(em)}_{ge}$ and $M^{(em, sf)}_{eg}$ are the matrix elements of the Hamiltonian terms~\eqref{eq:Horb} and \eqref{spin:phon} describing transitions between NC states without and with spin flip, respectively, involving emission of an optical phonon; correspondingly $M^{(abs)}_{ge}$ and $M^{(abs, sf)}_{eg}$ describe transitions involving absorption of an optical phonon. The specific expressions for the matrix elements are given in the Appendix. Equation~\eqref{eq:M1} is written as if there is only one excited state in a quantum dot. In general, the sum over all states $e$ should be carried out in the matrix element.

The relaxation rate reads
\begin{equation}
\label{eq:rate}
\Gamma_{\downarrow\uparrow} = \frac{2\pi}{\hbar} \sum_{\bm q_1,\bm q_2} |M_{\downarrow\uparrow}|^2 \Delta_\gamma(\hbar\Omega_1 - \hbar\Omega_2 - g \mu_B B).
\end{equation}
%\commV{Misha, this formula contain $M_{\downarrow\uparrow}$, while summation over excited states done in $M$ that is eq.(23). Is it correct that this formula should contain $M$ from eq.(23)? Shall we introduce summation over excited states already in (12) just for clarity?} \commentMisha{See addition above}
Here, for definiteness, $\Omega_1$ and $\Omega_2$ are the frequencies of the first and second phonon, the choice of signs corresponds to which of the phonons is emitted or absorbed, and
\begin{equation}
\label{broad}
\Delta_\gamma(x) = \frac{1}{\pi} \frac{\hbar\gamma}{x^2+(\hbar\gamma)^2}
\end{equation}
is the broadened $\delta$-function that accounts for the finite phonon lifetime $1/\gamma$ \cite{Trifonov2026,prokofiev2014phonon}. We assume that this broadening is large in comparison to the electron Zeeman splitting ($\hbar\gamma \gg |g \mu_B B|$) and to the relevant range of the optical-phonon dispersion. The latter is restricted by the condition $qa\ll 1$, otherwise the corresponding matrix elements are beyond the long-wavelength approximation and are suppressed. Furthermore, we assume that $\hbar\Omega_1 \approx \hbar\Omega_2 = \hbar\Omega_\text{LO} \ll \Delta_{eg}$.  We also assume that the temperature is large enough, so that $k_B T \gg |g \mu_B B|$, but $\hbar\Omega \ll k_B T$. %\commentMisha{To simplify notations, I suggest to omit the subscript LO in $\Omega_{\rm LO}$ hereafter.}\commV{It enters in the final formula and makes it more intuitive, so I would propose to keep it.} \addMisha{OK}

After some algebra, presented in the Appendix and restricting the summation over $\bm q_1$ and $\bm q_2$ in Eq.~\eqref{eq:rate} to the range of wavevectors satisfying the long-wavelength condition, we arrive at the following expression for the transition rate:
\begin{multline}
\label{eq:rate:3}
\Gamma_{\downarrow\uparrow} =
\frac{4C_\chi^2}{27\pi^2} \left( \frac{\varepsilon_\text{s}}{\varepsilon^*}\right)^2 \left(\frac{a}{a_B}\right)^2 \frac{\left(\hbar\Omega_\text{LO}\right)^{4}}{\hbar^2\gamma} \left(\frac{m \alpha_R}{e\hbar^2}\right)^2 \\
\times \exp{\left(-\frac{\hbar\Omega_\text{LO}}{k_B T}\right)},
\end{multline}
where $a_B=\hbar^2\varepsilon_\text{s}/(me^2)$ is the effective Bohr radius, $m$ is the electron effective mass, and $C_\chi$ is a numerical factor depending on the shape of the NC.
%\CommMS{Maybe it's useful to mention that calculations involve summation over all excited states of electrons in a NC, rather than just one as in the figure} \commentMisha{Done}

The Rashba constant $\alpha_R$ in perovskites can be evaluated following the approach for bulk semiconductors presented in Ref.~\cite{winklerSpinOrbitCoupling2003} (see also~\cite{andradaesilvaSpinsplitSubbandsMagnetooscillations1994}):
\begin{equation}
\label{alphaR}
\left|\frac{\alpha_R}{e}\right| =  \frac{P^2}{3E_g^{2}},
\end{equation}
where $E_g$ is the band gap in the perovskite (determined as the energy distance between the $\Gamma_7^-$ conduction and the $\Gamma_6^+$ valence band), $P = (\hbar/m_0)p$, $m_0$ is the free electron mass, and $p \equiv p_\parallel =p_\perp$ is the interband momentum matrix element (we assume the cubic phase for simplicity), see Refs.~\cite{Kirstein2022Uni,Nestoklon2023}.  The description of the band structure and the corresponding parameters are presented in Ref.~\cite{Kirstein2022Uni}. 

We see that Eq.~\eqref{eq:rate:3} resembles the temperature-dependent part of Eq.~\eqref{eq:act} used for fitting the experimental dependence with $E_\text{a} = \hbar \Omega_\text{LO}$. We note that this equation does not contain any dependence on the magnetic field. In our case the Zeeman splitting is much smaller than the phonon linewidth, and the energy conservation is basically fufilled regardless the details of the phonon spectra. In high fields, when the condition $\hbar\gamma \gg |g \mu_B B|$ is not fulfilled, the situation is more complex, and single acoustic phonons may become important. 

Let us use Eq.~\eqref{eq:rate:3} to estimate the prefactor $\gamma_\text{a}$ in Eq.~\eqref{eq:act}. For estimation purposes, we take $C_\chi = 1$, $\varepsilon_\text{s} = 30$,  $\varepsilon^* = 6$, $a/a_B \sim 1$, $P\approx 5$~eV\AA, $E_\text{g}\approx 1.7$~eV, $m = 0.18 m_0$ \cite{Kirstein2022Uni}, $\hbar\Omega_\text{LO} = 3.2$~meV, and $\gamma = 0.2$~ps$^{-1}$; the last two values are based on the experimental data of Ref.~\cite{Harkort2025,Trifonov2026}. With these values, we get an estimate for the exponent prefactor of $\gamma_\text{a} = 2$~$\mu$s$^{-1}$. Despite the approximations used and the rather arbitrary parameter choice, this estimate fits the experimental value within an order of magnitude. More importantly, the model reproduces the activation-like character of the $T_1$ temperature dependence, revealing the relevance of the two-LO-phonon mechanism of spin relaxation in perovskite NCs.

It is noteworthy, that although the resulting spin relaxation times are rather long and corresponing rates $\Gamma_{\downarrow\uparrow}$ are relatively low there is, in theory, a substantial room for suppressing $\Gamma_{\downarrow\uparrow}$ and enhancing the $T_1$. Indeed, the proposed mechansim relies on the fact that the energy conservation $\delta$-function, $\Delta_\gamma(\hbar\Omega_1 - \hbar\Omega_2 - g\mu_B B)$ is essentially broadened due to the uncertainty of the phonon energies, i.e., due to their finite lifetime. Understanding the mechanisms that limit the phonon lifetime and harnessing these mechansims may allow for control and prolongation of the charge carrier spin relaxation times.
 
\section{Conclusions}
We have measured ODMR resonances for electrons and holes in CsPbI$_3$ NCs and determined the key spin parameters: the $g$ factors and their spread, the longitudinal spin relaxation times $T_1$, and the effective nuclear fields resulting from the hyperfine interaction. The electron spin relaxation time $T_1$ is found to be twice longer than the $T_1$ for holes. We show that the electron $T_1$ strongly increases when the magnetic field is varied from 0 to 50~mT and saturates at higher fields. This behavior can be explained by the suppression of the impact of the time-varying effective Overhauser nuclear fields; from our data, we can determine the characteristic time of their variation $\tau_\text{c} \approx 56$~$\mu$s. At low laser powers and in strong magnetic fields, an electron spin relaxation time as long as 0.9~ms is found. We observe an activation-like enhancement of the spin relaxation with increasing temperature and describe it using a model based on a two-LO-phonon Raman process. This study suggests perovskite NCs in a glass matrix as a promising artificially created confined system for spintronics applications.

\section{Acknowledgments}
We acknowledge fruitful discussions with M.~O.~Nestoklon. V.V.B. and D.R.Y. acknowledge support of the Deutsche Forschungsgemeinschaft (Project YA 65/28-1, No. 527080192).
E.V.~Kolobkova and M.S.K. acknowledge support by the Saint Petersburg State University (Grant No. 125022803069-4). 

\begin{widetext}
\section{Appendix}
We consider a NC with the symmetry of a cuboid (rectangular parallelepiped) with $x$, $y$, and $z$ being its principal axes. The ground state is described by a fully symmetric function $g(\bm r)$. To begin with, let us consider only one intermediate excited state $e$ described by a function $e(\bm r)$ that transforms as the $y$ coordinate, as shown in Fig.~\ref{fig:T1}(c). For example, in a NC with infinitely high potential barriers, these functions are
\[
g(\bm r) =\mathcal N_g \cos{\left(\frac{\pi x}{a_x}\right)}\cos{\left(\frac{\pi y}{a_y}\right)}\cos{\left(\frac{\pi z}{a_z}\right)}, \quad e(\bm r) =\mathcal N_e \cos{\left(\frac{\pi x}{a_x}\right)}\sin{\left(\frac{2\pi y}{a_y}\right)}\cos{\left(\frac{\pi z}{a_z}\right)},
\]
where $a_x$, $a_y$, $a_z$ are the lengths of the cuboid's edges and $\mathcal N_g$, $\mathcal N_e$ are the normalization constants. A similar analysis for all other intermediate excited states is presented below.

The relevant matrix elements introduced in Eq.~\eqref{eq:M1} take the form:
\begin{subequations}
\begin{align}
M_{eg}^{(abs)}(\bm q)=-e\alpha_F\sqrt{n_{q}} \frac{q_y}{q} \langle e|y|g\rangle,\\ 
M_{eg}^{(em)}(\bm q)=-e\alpha_F\sqrt{n_{q}+1} \frac{q_y}{q} \langle e|y|g\rangle,\\
M_{ge}^{(abs)}(\bm q)=-e\alpha_F\sqrt{n_{q}} \frac{q_y}{q} \langle g|y|e\rangle = M_{eg}^{(abs)}(\bm q),\\ 
M_{ge}^{(em)}(\bm q)=-e\alpha_F\sqrt{n_{q}+1} \frac{q_y}{q} \langle g|y|e\rangle = M_{eg}^{(em)}(\bm q),
\end{align}
for the spin-conserving processes, and 
\begin{align}
M_{eg}^{(abs,sf)}(\bm q) = \alpha_F\alpha_R \sqrt{n_{q}} \frac{q_z}{q} \langle e|\hat{k}_y|g\rangle, \\
 M_{eg}^{(em,sf)}(\bm q) = \alpha_F\alpha_R \sqrt{n_{q}+1}\frac{q_z}{q} \langle e|\hat{k}_y|g\rangle,\\
M_{ge}^{(abs,sf)}(\bm q) = \alpha_F\alpha_R \sqrt{n_{q}} \frac{q_z}{q} \langle g|\hat{k}_y|e\rangle = - M_{eg}^{(abs,sf)}(\bm q),\\  
M_{ge}^{(em,sf)}(\bm q) = \alpha_F\alpha_R \sqrt{n_{q}+1}\frac{q_z}{q} \langle g|\hat{k}_y|e\rangle = -M_{eg}^{(em,sf)}(\bm q),
\end{align}
for the spin-flip processes. Here, we made use of the relations
\[
\langle e|y|g\rangle = \langle g|y|e\rangle, \quad \langle e|\hat{k}_y|g\rangle = - \langle g|\hat{k}_y|e\rangle,
\]
which are valid for the real wavefunctions of the initial and intermediate states. The filling factors $\sqrt{n_q}$ and $\sqrt{n_q+1}$ appear from the matrix elements of the annihilation and creation operators acting between the states with $n_q$ and $n_q \pm 1$ phonons. Strictly speaking, statistical averaging over the phonon distribution should be carried out in the expression for the transition rate. However, in the case where the phonon modes are uncorrelated and the result is linear in the occupancies of each mode, statistical averaging does not change the result.  %\commV{!!!}The occupancy factors in the matrix elements will be  $\sqrt{n_{q_1}(1+n_{q_2})}$.
\end{subequations}

For the total matrix element we get
\begin{subequations}
\label{M}
\begin{multline}
\label{M1}
M^{(1)}(\bm q_1,\bm q_2) = -\frac{M^{(em)}_{ge}(\bm q_2)M^{(abs, sf)}_{eg}(\bm q_1)}{\Delta_{eg}-\hbar\Omega_1}-\frac{M^{(abs)}_{ge}(\bm q_1)M^{(em, sf)}_{eg}(\bm q_2)}{\Delta_{eg}+\hbar\Omega_2}\\
=\frac{e\alpha_R \alpha_F^2}{q_1q_2} \sqrt{n_{q_1}(1+n_{q_2})}\langle e|y|g\rangle\langle e|\hat{k}_y|g\rangle \left(\frac{q_{2,y}q_{1,z}}{\Delta_{eg}-\hbar\Omega_1} + \frac{q_{1,y}q_{2,z}}{\Delta_{eg}+\hbar\Omega_2} \right),
\end{multline}
and
\begin{multline}
\label{M2}
M^{(2)}(\bm q_1,\bm q_2) = -\frac{M^{(em,sf)}_{ge}(\bm q_2)M^{(abs)}_{eg}(\bm q_1)}{\Delta_{eg}-\hbar\Omega_1}-\frac{M^{(abs,sf)}_{ge}(\bm q_1)M^{(em)}_{eg}(\bm q_2)}{\Delta_{eg}+\hbar\Omega_2}\\
=-\frac{e\alpha_R \alpha_F^2}{q_1q_2} \sqrt{n_{q_1}(1+n_{q_2})}\langle e|y|g\rangle\langle e|\hat{k}_y|g\rangle \left(\frac{q_{2,z}q_{1,y}}{\Delta_{eg}-\hbar\Omega_1} + \frac{q_{1,z}q_{2,y}}{\Delta_{eg}+\hbar\Omega_2} \right),
\end{multline}
\end{subequations}
Note that $M^{(1)}(\bm q_1, \bm q_2) = -M^{(2)}(\bm q_2,\bm q_1)$.

Taking into account that $\hbar\Omega_1 \approx \hbar\Omega_2 = \hbar\Omega_\text{LO} \ll \Delta_{eg}$ we expand the fractional expressions in Eqs.~\eqref{M} in a series in $\hbar\Omega_\text{LO}/\Delta_{eg}$ neglecting the difference between $\Omega_1$ and $\Omega_2$. As a result, we have
\begin{equation}
\label{M:1}
M_{\downarrow\uparrow} = M^{(1)}(\bm q_1, \bm q_2)  + M^{(2)}(\bm q_1, \bm q_2) = \frac{2e\alpha_R \alpha_F^2\hbar\Omega_\text{LO}}{q_1q_2\Delta_{eg}^2} \sqrt{n_{q_1}(1+n_{q_2})}\langle e|y|g\rangle\langle e|\hat{k}_y|g\rangle [\bm q_2 \times \bm q_1]_x.
\end{equation}
Since
\begin{equation}
\label{ky}
\langle e|\hat{k}_y|g\rangle = \frac{m}{\hbar} \langle e|dy/dt|g\rangle =\frac{m}{\hbar} \langle e|\frac{\mathrm i}{\hbar}[\mathcal H,y]|g\rangle = \frac{\mathrm i m}{\hbar^2} \Delta_{eg} \langle e|y|g\rangle,
\end{equation}
where $\mathcal H$ is the Hamiltonian describing the quantum-confinement effect, we obtain 
\begin{equation}
\label{M12}
M_{\downarrow\uparrow} = \frac{\mathrm i m}{\hbar^2} \frac{2e\alpha_R \alpha_F^2\hbar\Omega_\text{LO}}{q_1q_2\Delta_{eg}} \sqrt{n_{q_1}(1+n_{q_2})} |\langle e|y|g\rangle|^2 [\bm q_2 \times \bm q_1]_x.
\end{equation}

Introducing the static electronic polarizability tensor of the NC with the principal matrix elements ($i=x,y,z$~\cite{landauQuantumMechanicsNonRelativistic1977})
\begin{equation}
\label{chi}
\chi_{ii} = \sum_e \frac{2e^2 |\langle e|r_i|g\rangle|^2}{\Delta_{eg}}, \quad r_x=x,~r_y=y,~r_z=z,
\end{equation}
where the summation involves all excited states $e$, and taking into account the similar set of excited states that transform like the $x$ coordinate, we arrive at the following expression for the spin-dependent part of the matrix element
\begin{equation}
\hat{M}%_{\addV{\downarrow\uparrow}} 
=  \frac{\mathrm i m}{\hbar^2} (\alpha_R/e) \alpha_F^2 \sqrt{n_{q_1}(1+n_{q_2})} \hbar\Omega_\text{LO} \frac{\chi_{yy}[\bm q_2 \times \bm q_1]_x\hat{\sigma}_x + \chi_{xx}[\bm q_2 \times \bm q_1]_y\hat{\sigma}_y }{q_2q_1}.
\end{equation}

Thus, the transition rate [Eq.~\eqref{eq:rate}] reads
\begin{equation}
\label{eq:si:rate:1}
\Gamma_{\downarrow\uparrow} = \exp{\left(-\frac{\hbar\Omega_\text{LO}}{k_B T}\right)}\sum_{\bm q_1 \bm q_2} \frac{2\pi}{\hbar} \left|\frac{m}{\hbar^2} (\alpha_R/e) \alpha_F^2 \hbar\Omega_\text{LO} \right|^2 (\chi_{xx}^2+\chi_{yy}^2) \times \frac{1}{\pi\hbar\gamma} \times \frac{2}{3}.
\end{equation}
Here, the factor $2/3$ comes from angular averaging. The sum over $\bm q_{1,2}$ should be restricted to the range of wavevectors where the long-wavelength approximation holds, 
\[
\sum_{\bm q_1 \bm q_2} \ldots \sim \left(\frac{\mathcal V}{(2\pi)^3}\right)^2 (4\pi)^2 \int_0^{a^{-1}} q_1^2 dq_1\int_0^{a^{-1}} q_2^2 dq_2 = \frac{\mathcal V^2}{36\pi^4a^6}. 
\]
Combining all factors, we obtain
\begin{equation}
\label{rate:2}
\Gamma_{\downarrow\uparrow} = \frac{4}{27\pi^2}\exp{\left(-\frac{\hbar\Omega_\text{LO}}{k_B T}\right)} \frac{1}{\hbar^2\gamma} \left(\frac{m \alpha_R}{e\hbar^2}\right)^2 \frac{\left(\hbar\Omega_\text{LO}\right)^3}{\left(\varepsilon^*\right)^2} \frac{\chi_{xx}^2+\chi_{yy}^2}{a^6}.
\end{equation}
The susceptibility $\sqrt{\chi_{xx}^2+\chi_{yy}^2}=C_\chi\varepsilon_0 a^4/a_B$, where $a_B=\hbar^2\varepsilon_0/(me^2)$ and $C_\chi$ is the numerical factor depending on the shape of the wavefunctions. Hence, Eq.~\eqref{rate:2} can be rewritten in the form convenient for numerical estimates
\begin{equation}
\label{rate:3}
\Gamma_{\downarrow\uparrow} = \frac{4C_\chi^2}{27\pi^2} \left( \frac{\varepsilon_0}{\varepsilon^*}\right)^2 \left(\frac{a}{a_B}\right)^2 \exp{\left(-\frac{\hbar\Omega_\text{LO}}{k_B T}\right)} \frac{\left(\hbar\Omega_\text{LO}\right)^{4}}{\hbar^2\gamma} \left(\frac{m \alpha_R}{e\hbar^2}\right)^2.
\end{equation}

\end{widetext}


\begin{thebibliography}{57}%
\makeatletter
\providecommand \@ifxundefined [1]{%
 \@ifx{#1\undefined}
}%
\providecommand \@ifnum [1]{%
 \ifnum #1\expandafter \@firstoftwo
 \else \expandafter \@secondoftwo
 \fi
}%
\providecommand \@ifx [1]{%
 \ifx #1\expandafter \@firstoftwo
 \else \expandafter \@secondoftwo
 \fi
}%
\providecommand \natexlab [1]{#1}%
\providecommand \enquote  [1]{``#1''}%
\providecommand \bibnamefont  [1]{#1}%
\providecommand \bibfnamefont [1]{#1}%
\providecommand \citenamefont [1]{#1}%
\providecommand \href@noop [0]{\@secondoftwo}%
\providecommand \href [0]{\begingroup \@sanitize@url \@href}%
\providecommand \@href[1]{\@@startlink{#1}\@@href}%
\providecommand \@@href[1]{\endgroup#1\@@endlink}%
\providecommand \@sanitize@url [0]{\catcode `\\12\catcode `\$12\catcode
  `\&12\catcode `\#12\catcode `\^12\catcode `\_12\catcode `\%12\relax}%
\providecommand \@@startlink[1]{}%
\providecommand \@@endlink[0]{}%
\providecommand \url  [0]{\begingroup\@sanitize@url \@url }%
\providecommand \@url [1]{\endgroup\@href {#1}{\urlprefix }}%
\providecommand \urlprefix  [0]{URL }%
\providecommand \Eprint [0]{\href }%
\providecommand \doibase [0]{https://doi.org/}%
\providecommand \selectlanguage [0]{\@gobble}%
\providecommand \bibinfo  [0]{\@secondoftwo}%
\providecommand \bibfield  [0]{\@secondoftwo}%
\providecommand \translation [1]{[#1]}%
\providecommand \BibitemOpen [0]{}%
\providecommand \bibitemStop [0]{}%
\providecommand \bibitemNoStop [0]{.\EOS\space}%
\providecommand \EOS [0]{\spacefactor3000\relax}%
\providecommand \BibitemShut  [1]{\csname bibitem#1\endcsname}%
\let\auto@bib@innerbib\@empty
%</preamble>
\bibitem [{\citenamefont {Liu}\ \emph {et~al.}(2013)\citenamefont {Liu},
  \citenamefont {Johnston},\ and\ \citenamefont {Snaith}}]{Liu2013}%
  \BibitemOpen
  \bibfield  {author} {\bibinfo {author} {\bibfnamefont {M.}~\bibnamefont
  {Liu}}, \bibinfo {author} {\bibfnamefont {M.~B.}\ \bibnamefont {Johnston}},\
  and\ \bibinfo {author} {\bibfnamefont {H.~J.}\ \bibnamefont {Snaith}},\
  }\bibfield  {title} {\bibinfo {title} {{Efficient planar heterojunction
  perovskite solar cells by vapour deposition}},\ }\href
  {https://doi.org/10.1038/nature12509} {\bibfield  {journal} {\bibinfo
  {journal} {Nature}\ }\textbf {\bibinfo {volume} {501}},\ \bibinfo {pages}
  {395} (\bibinfo {year} {2013})}\BibitemShut {NoStop}%
\bibitem [{\citenamefont {Green}\ \emph {et~al.}(2014)\citenamefont {Green},
  \citenamefont {Ho-Baillie},\ and\ \citenamefont {Snaith}}]{Green2014}%
  \BibitemOpen
  \bibfield  {author} {\bibinfo {author} {\bibfnamefont {M.~A.}\ \bibnamefont
  {Green}}, \bibinfo {author} {\bibfnamefont {A.}~\bibnamefont {Ho-Baillie}},\
  and\ \bibinfo {author} {\bibfnamefont {H.~J.}\ \bibnamefont {Snaith}},\
  }\bibfield  {title} {\bibinfo {title} {{The emergence of perovskite solar
  cells}},\ }\href {https://doi.org/10.1038/nphoton.2014.134} {\bibfield
  {journal} {\bibinfo  {journal} {Nature Photonics}\ }\textbf {\bibinfo
  {volume} {8}},\ \bibinfo {pages} {506} (\bibinfo {year} {2014})}\BibitemShut
  {NoStop}%
\bibitem [{\citenamefont {Han}\ \emph {et~al.}(2025)\citenamefont {Han},
  \citenamefont {Park}, \citenamefont {Tan}, \citenamefont {Vaynzof},
  \citenamefont {Xue}, \citenamefont {Diau}, \citenamefont {Bawendi},
  \citenamefont {Lee},\ and\ \citenamefont {Jeon}}]{han2025perovskite}%
  \BibitemOpen
  \bibfield  {author} {\bibinfo {author} {\bibfnamefont {J.}~\bibnamefont
  {Han}}, \bibinfo {author} {\bibfnamefont {K.}~\bibnamefont {Park}}, \bibinfo
  {author} {\bibfnamefont {S.}~\bibnamefont {Tan}}, \bibinfo {author}
  {\bibfnamefont {Y.}~\bibnamefont {Vaynzof}}, \bibinfo {author} {\bibfnamefont
  {J.}~\bibnamefont {Xue}}, \bibinfo {author} {\bibfnamefont {E.~W.-G.}\
  \bibnamefont {Diau}}, \bibinfo {author} {\bibfnamefont {M.~G.}\ \bibnamefont
  {Bawendi}}, \bibinfo {author} {\bibfnamefont {J.-W.}\ \bibnamefont {Lee}},\
  and\ \bibinfo {author} {\bibfnamefont {I.}~\bibnamefont {Jeon}},\ }\bibfield
  {title} {\bibinfo {title} {Perovskite solar cells},\ }\href
  {https://doi.org/10.1038/s43586-024-00373-9} {\bibfield  {journal} {\bibinfo
  {journal} {Nature Reviews Methods Primers}\ }\textbf {\bibinfo {volume}
  {5}},\ \bibinfo {pages} {3} (\bibinfo {year} {2025})}\BibitemShut {NoStop}%
\bibitem [{\citenamefont {Stoumpos}\ \emph {et~al.}(2013)\citenamefont
  {Stoumpos}, \citenamefont {Malliakas}, \citenamefont {Peters}, \citenamefont
  {Liu}, \citenamefont {Sebastian}, \citenamefont {Im}, \citenamefont
  {Chasapis}, \citenamefont {Wibowo}, \citenamefont {Chung}, \citenamefont
  {Freeman}, \citenamefont {Wessels},\ and\ \citenamefont
  {Kanatzidis}}]{Stoumpos2013}%
  \BibitemOpen
  \bibfield  {author} {\bibinfo {author} {\bibfnamefont {C.~C.}\ \bibnamefont
  {Stoumpos}}, \bibinfo {author} {\bibfnamefont {C.~D.}\ \bibnamefont
  {Malliakas}}, \bibinfo {author} {\bibfnamefont {J.~A.}\ \bibnamefont
  {Peters}}, \bibinfo {author} {\bibfnamefont {Z.}~\bibnamefont {Liu}},
  \bibinfo {author} {\bibfnamefont {M.}~\bibnamefont {Sebastian}}, \bibinfo
  {author} {\bibfnamefont {J.}~\bibnamefont {Im}}, \bibinfo {author}
  {\bibfnamefont {T.~C.}\ \bibnamefont {Chasapis}}, \bibinfo {author}
  {\bibfnamefont {A.~C.}\ \bibnamefont {Wibowo}}, \bibinfo {author}
  {\bibfnamefont {D.~Y.}\ \bibnamefont {Chung}}, \bibinfo {author}
  {\bibfnamefont {A.~J.}\ \bibnamefont {Freeman}}, \bibinfo {author}
  {\bibfnamefont {B.~W.}\ \bibnamefont {Wessels}},\ and\ \bibinfo {author}
  {\bibfnamefont {M.~G.}\ \bibnamefont {Kanatzidis}},\ }\bibfield  {title}
  {\bibinfo {title} {{Crystal growth of the perovskite semiconductor
  CsPbBr$_3$: a new material for high-energy radiation detection}},\ }\href
  {https://doi.org/10.1021/cg400645t} {\bibfield  {journal} {\bibinfo
  {journal} {Crystal Growth and Design}\ }\textbf {\bibinfo {volume} {13}},\
  \bibinfo {pages} {2722} (\bibinfo {year} {2013})}\BibitemShut {NoStop}%
\bibitem [{\citenamefont {Yakunin}\ \emph {et~al.}(2015)\citenamefont
  {Yakunin}, \citenamefont {Sytnyk}, \citenamefont {Kriegner}, \citenamefont
  {Shrestha}, \citenamefont {Richter}, \citenamefont {Matt}, \citenamefont
  {Azimi}, \citenamefont {Brabec}, \citenamefont {Stangl}, \citenamefont
  {Kovalenko},\ and\ \citenamefont {Heiss}}]{Yakunin2015}%
  \BibitemOpen
  \bibfield  {author} {\bibinfo {author} {\bibfnamefont {S.}~\bibnamefont
  {Yakunin}}, \bibinfo {author} {\bibfnamefont {M.}~\bibnamefont {Sytnyk}},
  \bibinfo {author} {\bibfnamefont {D.}~\bibnamefont {Kriegner}}, \bibinfo
  {author} {\bibfnamefont {S.}~\bibnamefont {Shrestha}}, \bibinfo {author}
  {\bibfnamefont {M.}~\bibnamefont {Richter}}, \bibinfo {author} {\bibfnamefont
  {G.~J.}\ \bibnamefont {Matt}}, \bibinfo {author} {\bibfnamefont
  {H.}~\bibnamefont {Azimi}}, \bibinfo {author} {\bibfnamefont {C.~J.}\
  \bibnamefont {Brabec}}, \bibinfo {author} {\bibfnamefont {J.}~\bibnamefont
  {Stangl}}, \bibinfo {author} {\bibfnamefont {M.~V.}\ \bibnamefont
  {Kovalenko}},\ and\ \bibinfo {author} {\bibfnamefont {W.}~\bibnamefont
  {Heiss}},\ }\bibfield  {title} {\bibinfo {title} {{Detection of X-ray photons
  by solution-processed lead halide perovskites}},\ }\href
  {https://doi.org/10.1038/nphoton.2015.82} {\bibfield  {journal} {\bibinfo
  {journal} {Nature Photonics}\ }\textbf {\bibinfo {volume} {9}},\ \bibinfo
  {pages} {444} (\bibinfo {year} {2015})}\BibitemShut {NoStop}%
\bibitem [{\citenamefont {Cao}\ \emph {et~al.}(2018)\citenamefont {Cao},
  \citenamefont {Wang}, \citenamefont {Tian}, \citenamefont {Guo},
  \citenamefont {Wei}, \citenamefont {Chen}, \citenamefont {Miao},
  \citenamefont {Zou}, \citenamefont {Pan}, \citenamefont {He}, \citenamefont
  {Cao}, \citenamefont {Ke}, \citenamefont {Xu}, \citenamefont {Wang},
  \citenamefont {Yang}, \citenamefont {Du}, \citenamefont {Fu}, \citenamefont
  {Kong}, \citenamefont {Dai}, \citenamefont {Jin}, \citenamefont {Li},
  \citenamefont {Li}, \citenamefont {Peng}, \citenamefont {Wang},\ and\
  \citenamefont {Huang}}]{Cao2018}%
  \BibitemOpen
  \bibfield  {author} {\bibinfo {author} {\bibfnamefont {Y.}~\bibnamefont
  {Cao}}, \bibinfo {author} {\bibfnamefont {N.}~\bibnamefont {Wang}}, \bibinfo
  {author} {\bibfnamefont {H.}~\bibnamefont {Tian}}, \bibinfo {author}
  {\bibfnamefont {J.}~\bibnamefont {Guo}}, \bibinfo {author} {\bibfnamefont
  {Y.}~\bibnamefont {Wei}}, \bibinfo {author} {\bibfnamefont {H.}~\bibnamefont
  {Chen}}, \bibinfo {author} {\bibfnamefont {Y.}~\bibnamefont {Miao}}, \bibinfo
  {author} {\bibfnamefont {W.}~\bibnamefont {Zou}}, \bibinfo {author}
  {\bibfnamefont {K.}~\bibnamefont {Pan}}, \bibinfo {author} {\bibfnamefont
  {Y.}~\bibnamefont {He}}, \bibinfo {author} {\bibfnamefont {H.}~\bibnamefont
  {Cao}}, \bibinfo {author} {\bibfnamefont {Y.}~\bibnamefont {Ke}}, \bibinfo
  {author} {\bibfnamefont {M.}~\bibnamefont {Xu}}, \bibinfo {author}
  {\bibfnamefont {Y.}~\bibnamefont {Wang}}, \bibinfo {author} {\bibfnamefont
  {M.}~\bibnamefont {Yang}}, \bibinfo {author} {\bibfnamefont {K.}~\bibnamefont
  {Du}}, \bibinfo {author} {\bibfnamefont {Z.}~\bibnamefont {Fu}}, \bibinfo
  {author} {\bibfnamefont {D.}~\bibnamefont {Kong}}, \bibinfo {author}
  {\bibfnamefont {D.}~\bibnamefont {Dai}}, \bibinfo {author} {\bibfnamefont
  {Y.}~\bibnamefont {Jin}}, \bibinfo {author} {\bibfnamefont {G.}~\bibnamefont
  {Li}}, \bibinfo {author} {\bibfnamefont {H.}~\bibnamefont {Li}}, \bibinfo
  {author} {\bibfnamefont {Q.}~\bibnamefont {Peng}}, \bibinfo {author}
  {\bibfnamefont {J.}~\bibnamefont {Wang}},\ and\ \bibinfo {author}
  {\bibfnamefont {W.}~\bibnamefont {Huang}},\ }\bibfield  {title} {\bibinfo
  {title} {{Perovskite light-emitting diodes based on spontaneously formed
  submicrometre-scale structures}},\ }\href
  {https://doi.org/10.1038/s41586-018-0576-2} {\bibfield  {journal} {\bibinfo
  {journal} {Nature}\ }\textbf {\bibinfo {volume} {562}},\ \bibinfo {pages}
  {249} (\bibinfo {year} {2018})}\BibitemShut {NoStop}%
\bibitem [{\citenamefont {Liu}\ \emph {et~al.}(2021)\citenamefont {Liu},
  \citenamefont {Xu}, \citenamefont {Bai}, \citenamefont {Jin}, \citenamefont
  {Wang}, \citenamefont {Friend},\ and\ \citenamefont {Gao}}]{Liu2021}%
  \BibitemOpen
  \bibfield  {author} {\bibinfo {author} {\bibfnamefont {X.-K.}\ \bibnamefont
  {Liu}}, \bibinfo {author} {\bibfnamefont {W.}~\bibnamefont {Xu}}, \bibinfo
  {author} {\bibfnamefont {S.}~\bibnamefont {Bai}}, \bibinfo {author}
  {\bibfnamefont {Y.}~\bibnamefont {Jin}}, \bibinfo {author} {\bibfnamefont
  {J.}~\bibnamefont {Wang}}, \bibinfo {author} {\bibfnamefont {R.~H.}\
  \bibnamefont {Friend}},\ and\ \bibinfo {author} {\bibfnamefont
  {F.}~\bibnamefont {Gao}},\ }\bibfield  {title} {\bibinfo {title} {{Metal
  halide perovskites for light-emitting diodes}},\ }\href
  {https://doi.org/10.1038/s41563-020-0784-7} {\bibfield  {journal} {\bibinfo
  {journal} {Nature Materials}\ }\textbf {\bibinfo {volume} {20}},\ \bibinfo
  {pages} {10} (\bibinfo {year} {2021})}\BibitemShut {NoStop}%
\bibitem [{\citenamefont {Kovalenko}\ \emph {et~al.}(2017)\citenamefont
  {Kovalenko}, \citenamefont {Protesescu},\ and\ \citenamefont
  {Bodnarchuk}}]{Kovalenko2017}%
  \BibitemOpen
  \bibfield  {author} {\bibinfo {author} {\bibfnamefont {M.~V.}\ \bibnamefont
  {Kovalenko}}, \bibinfo {author} {\bibfnamefont {L.}~\bibnamefont
  {Protesescu}},\ and\ \bibinfo {author} {\bibfnamefont {M.~I.}\ \bibnamefont
  {Bodnarchuk}},\ }\bibfield  {title} {\bibinfo {title} {{Properties and
  potential optoelectronic applications of lead halide perovskite
  nanocrystals}},\ }\href {https://doi.org/10.1126/science.aam7093} {\bibfield
  {journal} {\bibinfo  {journal} {Science}\ }\textbf {\bibinfo {volume}
  {358}},\ \bibinfo {pages} {745} (\bibinfo {year} {2017})}\BibitemShut
  {NoStop}%
\bibitem [{\citenamefont {Utzat}\ \emph {et~al.}(2019)\citenamefont {Utzat},
  \citenamefont {Sun}, \citenamefont {Kaplan}, \citenamefont {Krieg},
  \citenamefont {Ginterseder}, \citenamefont {Spokoyny}, \citenamefont {Klein},
  \citenamefont {Shulenberger}, \citenamefont {Perkinson}, \citenamefont
  {Kovalenko},\ and\ \citenamefont {Bawendi}}]{Utzat2019}%
  \BibitemOpen
  \bibfield  {author} {\bibinfo {author} {\bibfnamefont {H.}~\bibnamefont
  {Utzat}}, \bibinfo {author} {\bibfnamefont {W.}~\bibnamefont {Sun}}, \bibinfo
  {author} {\bibfnamefont {A.~E.~K.}\ \bibnamefont {Kaplan}}, \bibinfo {author}
  {\bibfnamefont {F.}~\bibnamefont {Krieg}}, \bibinfo {author} {\bibfnamefont
  {M.}~\bibnamefont {Ginterseder}}, \bibinfo {author} {\bibfnamefont
  {B.}~\bibnamefont {Spokoyny}}, \bibinfo {author} {\bibfnamefont {N.~D.}\
  \bibnamefont {Klein}}, \bibinfo {author} {\bibfnamefont {K.~E.}\ \bibnamefont
  {Shulenberger}}, \bibinfo {author} {\bibfnamefont {C.~F.}\ \bibnamefont
  {Perkinson}}, \bibinfo {author} {\bibfnamefont {M.~V.}\ \bibnamefont
  {Kovalenko}},\ and\ \bibinfo {author} {\bibfnamefont {M.~G.}\ \bibnamefont
  {Bawendi}},\ }\bibfield  {title} {\bibinfo {title} {{Coherent single-photon
  emission from colloidal lead halide perovskite quantum dots}},\ }\href
  {https://doi.org/10.1126/science.aau7392} {\bibfield  {journal} {\bibinfo
  {journal} {Science}\ }\textbf {\bibinfo {volume} {363}},\ \bibinfo {pages}
  {1068} (\bibinfo {year} {2019})}\BibitemShut {NoStop}%
\bibitem [{\citenamefont {Protesescu}\ \emph {et~al.}(2015)\citenamefont
  {Protesescu}, \citenamefont {Yakunin}, \citenamefont {Bodnarchuk},
  \citenamefont {Krieg}, \citenamefont {Caputo}, \citenamefont {Hendon},
  \citenamefont {Yang}, \citenamefont {Walsh},\ and\ \citenamefont
  {Kovalenko}}]{Protesescu2015}%
  \BibitemOpen
  \bibfield  {author} {\bibinfo {author} {\bibfnamefont {L.}~\bibnamefont
  {Protesescu}}, \bibinfo {author} {\bibfnamefont {S.}~\bibnamefont {Yakunin}},
  \bibinfo {author} {\bibfnamefont {M.~I.}\ \bibnamefont {Bodnarchuk}},
  \bibinfo {author} {\bibfnamefont {F.}~\bibnamefont {Krieg}}, \bibinfo
  {author} {\bibfnamefont {R.}~\bibnamefont {Caputo}}, \bibinfo {author}
  {\bibfnamefont {C.~H.}\ \bibnamefont {Hendon}}, \bibinfo {author}
  {\bibfnamefont {R.~X.}\ \bibnamefont {Yang}}, \bibinfo {author}
  {\bibfnamefont {A.}~\bibnamefont {Walsh}},\ and\ \bibinfo {author}
  {\bibfnamefont {M.~V.}\ \bibnamefont {Kovalenko}},\ }\bibfield  {title}
  {\bibinfo {title} {{Nanocrystals of Cesium lead halide perovskites (CsPbX$_3$
  , X = Cl, Br, and I): novel optoelectronic materials showing bright emission
  with wide color gamut}},\ }\href {https://doi.org/10.1021/nl5048779}
  {\bibfield  {journal} {\bibinfo  {journal} {Nano Letters}\ }\textbf {\bibinfo
  {volume} {15}},\ \bibinfo {pages} {3692} (\bibinfo {year}
  {2015})}\BibitemShut {NoStop}%
\bibitem [{\citenamefont {Kolobkova}\ \emph {et~al.}(2021)\citenamefont
  {Kolobkova}, \citenamefont {Kuznetsova},\ and\ \citenamefont
  {Nikonorov}}]{Kolobkova2021}%
  \BibitemOpen
  \bibfield  {author} {\bibinfo {author} {\bibfnamefont {E.~V.}\ \bibnamefont
  {Kolobkova}}, \bibinfo {author} {\bibfnamefont {M.~S.}\ \bibnamefont
  {Kuznetsova}},\ and\ \bibinfo {author} {\bibfnamefont {N.~V.}\ \bibnamefont
  {Nikonorov}},\ }\bibfield  {title} {\bibinfo {title} {{Perovskite CsPbX$_3$
  (X=Cl, Br, I) nanocrystals in fluorophosphate glasses}},\ }\href
  {https://doi.org/10.1016/j.jnoncrysol.2021.120811} {\bibfield  {journal}
  {\bibinfo  {journal} {Journal of Non-Crystalline Solids}\ }\textbf {\bibinfo
  {volume} {563}},\ \bibinfo {pages} {120811} (\bibinfo {year}
  {2021})}\BibitemShut {NoStop}%
\bibitem [{\citenamefont {Nestoklon}\ \emph {et~al.}(2018)\citenamefont
  {Nestoklon}, \citenamefont {Goupalov}, \citenamefont {Dzhioev}, \citenamefont
  {Ken}, \citenamefont {Korenev}, \citenamefont {Kusrayev}, \citenamefont
  {Sapega}, \citenamefont {de~Weerd}, \citenamefont {Gomez}, \citenamefont
  {Gregorkiewicz}, \citenamefont {Lin}, \citenamefont {Suenaga}, \citenamefont
  {Fujiwara}, \citenamefont {Matyushkin},\ and\ \citenamefont
  {Yassievich}}]{Nestoklon2018}%
  \BibitemOpen
  \bibfield  {author} {\bibinfo {author} {\bibfnamefont {M.~O.}\ \bibnamefont
  {Nestoklon}}, \bibinfo {author} {\bibfnamefont {S.~V.}\ \bibnamefont
  {Goupalov}}, \bibinfo {author} {\bibfnamefont {R.~I.}\ \bibnamefont
  {Dzhioev}}, \bibinfo {author} {\bibfnamefont {O.~S.}\ \bibnamefont {Ken}},
  \bibinfo {author} {\bibfnamefont {V.~L.}\ \bibnamefont {Korenev}}, \bibinfo
  {author} {\bibfnamefont {Y.~G.}\ \bibnamefont {Kusrayev}}, \bibinfo {author}
  {\bibfnamefont {V.~F.}\ \bibnamefont {Sapega}}, \bibinfo {author}
  {\bibfnamefont {C.}~\bibnamefont {de~Weerd}}, \bibinfo {author}
  {\bibfnamefont {L.}~\bibnamefont {Gomez}}, \bibinfo {author} {\bibfnamefont
  {T.}~\bibnamefont {Gregorkiewicz}}, \bibinfo {author} {\bibfnamefont
  {J.}~\bibnamefont {Lin}}, \bibinfo {author} {\bibfnamefont {K.}~\bibnamefont
  {Suenaga}}, \bibinfo {author} {\bibfnamefont {Y.}~\bibnamefont {Fujiwara}},
  \bibinfo {author} {\bibfnamefont {L.~B.}\ \bibnamefont {Matyushkin}},\ and\
  \bibinfo {author} {\bibfnamefont {I.~N.}\ \bibnamefont {Yassievich}},\
  }\bibfield  {title} {\bibinfo {title} {{Optical orientation and alignment of
  excitons in ensembles of inorganic perovskite nanocrystals}},\ }\href
  {https://doi.org/10.1103/PhysRevB.97.235304} {\bibfield  {journal} {\bibinfo
  {journal} {Physical Review B}\ }\textbf {\bibinfo {volume} {97}},\ \bibinfo
  {pages} {235304} (\bibinfo {year} {2018})}\BibitemShut {NoStop}%
\bibitem [{\citenamefont {Odenthal}\ \emph {et~al.}(2017)\citenamefont
  {Odenthal}, \citenamefont {Talmadge}, \citenamefont {Gundlach}, \citenamefont
  {Wang}, \citenamefont {Zhang}, \citenamefont {Sun}, \citenamefont {Yu},
  \citenamefont {{Valy Vardeny}},\ and\ \citenamefont {Li}}]{Odenthal2017}%
  \BibitemOpen
  \bibfield  {author} {\bibinfo {author} {\bibfnamefont {P.}~\bibnamefont
  {Odenthal}}, \bibinfo {author} {\bibfnamefont {W.}~\bibnamefont {Talmadge}},
  \bibinfo {author} {\bibfnamefont {N.}~\bibnamefont {Gundlach}}, \bibinfo
  {author} {\bibfnamefont {R.}~\bibnamefont {Wang}}, \bibinfo {author}
  {\bibfnamefont {C.}~\bibnamefont {Zhang}}, \bibinfo {author} {\bibfnamefont
  {D.}~\bibnamefont {Sun}}, \bibinfo {author} {\bibfnamefont {Z.-G.}\
  \bibnamefont {Yu}}, \bibinfo {author} {\bibfnamefont {Z.}~\bibnamefont {{Valy
  Vardeny}}},\ and\ \bibinfo {author} {\bibfnamefont {Y.~S.}\ \bibnamefont
  {Li}},\ }\bibfield  {title} {\bibinfo {title} {{Spin-polarized exciton
  quantum beating in hybrid organic-inorganic perovskites}},\ }\href
  {https://doi.org/10.1038/nphys4145} {\bibfield  {journal} {\bibinfo
  {journal} {Nature Physics}\ }\textbf {\bibinfo {volume} {13}},\ \bibinfo
  {pages} {894} (\bibinfo {year} {2017})}\BibitemShut {NoStop}%
\bibitem [{\citenamefont {Belykh}\ \emph {et~al.}(2019)\citenamefont {Belykh},
  \citenamefont {Yakovlev}, \citenamefont {Glazov}, \citenamefont {Grigoryev},
  \citenamefont {Hussain}, \citenamefont {Rautert}, \citenamefont {Dirin},
  \citenamefont {Kovalenko},\ and\ \citenamefont {Bayer}}]{Belykh2019}%
  \BibitemOpen
  \bibfield  {author} {\bibinfo {author} {\bibfnamefont {V.~V.}\ \bibnamefont
  {Belykh}}, \bibinfo {author} {\bibfnamefont {D.~R.}\ \bibnamefont
  {Yakovlev}}, \bibinfo {author} {\bibfnamefont {M.~M.}\ \bibnamefont
  {Glazov}}, \bibinfo {author} {\bibfnamefont {P.~S.}\ \bibnamefont
  {Grigoryev}}, \bibinfo {author} {\bibfnamefont {M.}~\bibnamefont {Hussain}},
  \bibinfo {author} {\bibfnamefont {J.}~\bibnamefont {Rautert}}, \bibinfo
  {author} {\bibfnamefont {D.~N.}\ \bibnamefont {Dirin}}, \bibinfo {author}
  {\bibfnamefont {M.~V.}\ \bibnamefont {Kovalenko}},\ and\ \bibinfo {author}
  {\bibfnamefont {M.}~\bibnamefont {Bayer}},\ }\bibfield  {title} {\bibinfo
  {title} {{Coherent spin dynamics of electrons and holes in CsPbBr$_3$
  perovskite crystals}},\ }\href {https://doi.org/10.1038/s41467-019-08625-z}
  {\bibfield  {journal} {\bibinfo  {journal} {Nature Communications}\ }\textbf
  {\bibinfo {volume} {10}},\ \bibinfo {pages} {673} (\bibinfo {year}
  {2019})}\BibitemShut {NoStop}%
\bibitem [{\citenamefont {Crane}\ \emph {et~al.}(2020)\citenamefont {Crane},
  \citenamefont {Jacoby}, \citenamefont {Cohen}, \citenamefont {Huang},
  \citenamefont {Luscombe},\ and\ \citenamefont {Gamelin}}]{Crane2020}%
  \BibitemOpen
  \bibfield  {author} {\bibinfo {author} {\bibfnamefont {M.~J.}\ \bibnamefont
  {Crane}}, \bibinfo {author} {\bibfnamefont {L.~M.}\ \bibnamefont {Jacoby}},
  \bibinfo {author} {\bibfnamefont {T.~A.}\ \bibnamefont {Cohen}}, \bibinfo
  {author} {\bibfnamefont {Y.}~\bibnamefont {Huang}}, \bibinfo {author}
  {\bibfnamefont {C.~K.}\ \bibnamefont {Luscombe}},\ and\ \bibinfo {author}
  {\bibfnamefont {D.~R.}\ \bibnamefont {Gamelin}},\ }\bibfield  {title}
  {\bibinfo {title} {{Coherent Spin Precession and Lifetime-Limited Spin
  Dephasing in CsPbBr$_3$ Perovskite Nanocrystals}},\ }\href
  {https://doi.org/10.1021/acs.nanolett.0c03329} {\bibfield  {journal}
  {\bibinfo  {journal} {Nano Letters}\ }\textbf {\bibinfo {volume} {20}},\
  \bibinfo {pages} {8626} (\bibinfo {year} {2020})}\BibitemShut {NoStop}%
\bibitem [{\citenamefont {Grigoryev}\ \emph {et~al.}(2021)\citenamefont
  {Grigoryev}, \citenamefont {Belykh}, \citenamefont {Yakovlev}, \citenamefont
  {Lhuillier},\ and\ \citenamefont {Bayer}}]{Grigoryev2021}%
  \BibitemOpen
  \bibfield  {author} {\bibinfo {author} {\bibfnamefont {P.~S.}\ \bibnamefont
  {Grigoryev}}, \bibinfo {author} {\bibfnamefont {V.~V.}\ \bibnamefont
  {Belykh}}, \bibinfo {author} {\bibfnamefont {D.~R.}\ \bibnamefont
  {Yakovlev}}, \bibinfo {author} {\bibfnamefont {E.}~\bibnamefont
  {Lhuillier}},\ and\ \bibinfo {author} {\bibfnamefont {M.}~\bibnamefont
  {Bayer}},\ }\bibfield  {title} {\bibinfo {title} {{Coherent spin dynamics of
  electrons and holes in CsPbBr$_3$ colloidal nanocrystals}},\ }\href
  {https://doi.org/10.1021/acs.nanolett.1c03292} {\bibfield  {journal}
  {\bibinfo  {journal} {Nano Letters}\ }\textbf {\bibinfo {volume} {21}},\
  \bibinfo {pages} {8481} (\bibinfo {year} {2021})}\BibitemShut {NoStop}%
\bibitem [{\citenamefont {Kirstein}\ \emph
  {et~al.}(2022{\natexlab{a}})\citenamefont {Kirstein}, \citenamefont
  {Yakovlev}, \citenamefont {Glazov}, \citenamefont {Zhukov}, \citenamefont
  {Kudlacik}, \citenamefont {Kalitukha}, \citenamefont {Sapega}, \citenamefont
  {Dimitriev}, \citenamefont {Semina}, \citenamefont {Nestoklon}, \citenamefont
  {Ivchenko}, \citenamefont {Kopteva}, \citenamefont {Dirin}, \citenamefont
  {Nazarenko}, \citenamefont {Kovalenko}, \citenamefont {Baumann},
  \citenamefont {H{\"{o}}cker}, \citenamefont {Dyakonov},\ and\ \citenamefont
  {Bayer}}]{Kirstein2022Uni}%
  \BibitemOpen
  \bibfield  {author} {\bibinfo {author} {\bibfnamefont {E.}~\bibnamefont
  {Kirstein}}, \bibinfo {author} {\bibfnamefont {D.~R.}\ \bibnamefont
  {Yakovlev}}, \bibinfo {author} {\bibfnamefont {M.~M.}\ \bibnamefont
  {Glazov}}, \bibinfo {author} {\bibfnamefont {E.~A.}\ \bibnamefont {Zhukov}},
  \bibinfo {author} {\bibfnamefont {D.}~\bibnamefont {Kudlacik}}, \bibinfo
  {author} {\bibfnamefont {I.~V.}\ \bibnamefont {Kalitukha}}, \bibinfo {author}
  {\bibfnamefont {V.~F.}\ \bibnamefont {Sapega}}, \bibinfo {author}
  {\bibfnamefont {G.~S.}\ \bibnamefont {Dimitriev}}, \bibinfo {author}
  {\bibfnamefont {M.~A.}\ \bibnamefont {Semina}}, \bibinfo {author}
  {\bibfnamefont {M.~O.}\ \bibnamefont {Nestoklon}}, \bibinfo {author}
  {\bibfnamefont {E.~L.}\ \bibnamefont {Ivchenko}}, \bibinfo {author}
  {\bibfnamefont {N.~E.}\ \bibnamefont {Kopteva}}, \bibinfo {author}
  {\bibfnamefont {D.~N.}\ \bibnamefont {Dirin}}, \bibinfo {author}
  {\bibfnamefont {O.}~\bibnamefont {Nazarenko}}, \bibinfo {author}
  {\bibfnamefont {M.~V.}\ \bibnamefont {Kovalenko}}, \bibinfo {author}
  {\bibfnamefont {A.}~\bibnamefont {Baumann}}, \bibinfo {author} {\bibfnamefont
  {J.}~\bibnamefont {H{\"{o}}cker}}, \bibinfo {author} {\bibfnamefont
  {V.}~\bibnamefont {Dyakonov}},\ and\ \bibinfo {author} {\bibfnamefont
  {M.}~\bibnamefont {Bayer}},\ }\bibfield  {title} {\bibinfo {title} {{The
  Land{\'{e}} factors of electrons and holes in lead halide perovskites:
  universal dependence on the band gap}},\ }\href
  {https://doi.org/10.1038/s41467-022-30701-0} {\bibfield  {journal} {\bibinfo
  {journal} {Nature Communications}\ }\textbf {\bibinfo {volume} {13}},\
  \bibinfo {pages} {3062} (\bibinfo {year} {2022}{\natexlab{a}})}\BibitemShut
  {NoStop}%
\bibitem [{\citenamefont {Nestoklon}\ \emph
  {et~al.}(2023{\natexlab{a}})\citenamefont {Nestoklon}, \citenamefont
  {Kirstein}, \citenamefont {Yakovlev}, \citenamefont {Zhukov}, \citenamefont
  {Glazov}, \citenamefont {Semina}, \citenamefont {Ivchenko}, \citenamefont
  {Kolobkova}, \citenamefont {Kuznetsova},\ and\ \citenamefont
  {Bayer}}]{Nestoklon2023}%
  \BibitemOpen
  \bibfield  {author} {\bibinfo {author} {\bibfnamefont {M.~O.}\ \bibnamefont
  {Nestoklon}}, \bibinfo {author} {\bibfnamefont {E.}~\bibnamefont {Kirstein}},
  \bibinfo {author} {\bibfnamefont {D.~R.}\ \bibnamefont {Yakovlev}}, \bibinfo
  {author} {\bibfnamefont {E.~A.}\ \bibnamefont {Zhukov}}, \bibinfo {author}
  {\bibfnamefont {M.~M.}\ \bibnamefont {Glazov}}, \bibinfo {author}
  {\bibfnamefont {M.~A.}\ \bibnamefont {Semina}}, \bibinfo {author}
  {\bibfnamefont {E.~L.}\ \bibnamefont {Ivchenko}}, \bibinfo {author}
  {\bibfnamefont {E.~V.}\ \bibnamefont {Kolobkova}}, \bibinfo {author}
  {\bibfnamefont {M.~S.}\ \bibnamefont {Kuznetsova}},\ and\ \bibinfo {author}
  {\bibfnamefont {M.}~\bibnamefont {Bayer}},\ }\bibfield  {title} {\bibinfo
  {title} {{Tailoring the electron and hole land{\'{e}} factors in lead halide
  perovskite nanocrystals by quantum confinement and halide exchange}},\ }\href
  {https://doi.org/10.1021/acs.nanolett.3c02349} {\bibfield  {journal}
  {\bibinfo  {journal} {Nano Letters}\ }\textbf {\bibinfo {volume} {23}},\
  \bibinfo {pages} {8218} (\bibinfo {year} {2023}{\natexlab{a}})}\BibitemShut
  {NoStop}%
\bibitem [{\citenamefont {Kirstein}\ \emph {et~al.}(2025)\citenamefont
  {Kirstein}, \citenamefont {Yakovlev}, \citenamefont {Zhukov}, \citenamefont
  {Kopteva}, \citenamefont {Turedi}, \citenamefont {Kovalenko},\ and\
  \citenamefont {Bayer}}]{Kirstein2025RSA}%
  \BibitemOpen
  \bibfield  {author} {\bibinfo {author} {\bibfnamefont {E.}~\bibnamefont
  {Kirstein}}, \bibinfo {author} {\bibfnamefont {D.~R.}\ \bibnamefont
  {Yakovlev}}, \bibinfo {author} {\bibfnamefont {E.~A.}\ \bibnamefont
  {Zhukov}}, \bibinfo {author} {\bibfnamefont {N.~E.}\ \bibnamefont {Kopteva}},
  \bibinfo {author} {\bibfnamefont {B.}~\bibnamefont {Turedi}}, \bibinfo
  {author} {\bibfnamefont {M.~V.}\ \bibnamefont {Kovalenko}},\ and\ \bibinfo
  {author} {\bibfnamefont {M.}~\bibnamefont {Bayer}},\ }\bibfield  {title}
  {\bibinfo {title} {{Resonant spin amplification and accumulation in MAPbI$_3$
  single crystals}},\ }\href {https://doi.org/10.1002/advs.202502735}
  {\bibfield  {journal} {\bibinfo  {journal} {Advanced Science}\ }\textbf
  {\bibinfo {volume} {12}},\ \bibinfo {pages} {2502735} (\bibinfo {year}
  {2025})}\BibitemShut {NoStop}%
\bibitem [{\citenamefont {Meliakov}\ \emph {et~al.}(2025)\citenamefont
  {Meliakov}, \citenamefont {Zhukov}, \citenamefont {Belykh}, \citenamefont
  {Nestoklon}, \citenamefont {Kolobkova}, \citenamefont {Kuznetsova},
  \citenamefont {Bayer},\ and\ \citenamefont {Yakovlev}}]{Meliakov2025IE}%
  \BibitemOpen
  \bibfield  {author} {\bibinfo {author} {\bibfnamefont {S.~R.}\ \bibnamefont
  {Meliakov}}, \bibinfo {author} {\bibfnamefont {E.~A.}\ \bibnamefont
  {Zhukov}}, \bibinfo {author} {\bibfnamefont {V.~V.}\ \bibnamefont {Belykh}},
  \bibinfo {author} {\bibfnamefont {M.~O.}\ \bibnamefont {Nestoklon}}, \bibinfo
  {author} {\bibfnamefont {E.~V.}\ \bibnamefont {Kolobkova}}, \bibinfo {author}
  {\bibfnamefont {M.~S.}\ \bibnamefont {Kuznetsova}}, \bibinfo {author}
  {\bibfnamefont {M.}~\bibnamefont {Bayer}},\ and\ \bibinfo {author}
  {\bibfnamefont {D.~R.}\ \bibnamefont {Yakovlev}},\ }\bibfield  {title}
  {\bibinfo {title} {{Land{\'{e}} $g$-factors of electrons and holes strongly
  confined in CsPbI$_3$ perovskite nanocrystals in glass}},\ }\href
  {https://doi.org/10.1039/D4NR04602A} {\bibfield  {journal} {\bibinfo
  {journal} {Nanoscale}\ }\textbf {\bibinfo {volume} {17}},\ \bibinfo {pages}
  {6522} (\bibinfo {year} {2025})}\BibitemShut {NoStop}%
\bibitem [{\citenamefont {Zhu}\ \emph {et~al.}(2024)\citenamefont {Zhu},
  \citenamefont {Li}, \citenamefont {Lin}, \citenamefont {Han},\ and\
  \citenamefont {Wu}}]{zhu2024coherent}%
  \BibitemOpen
  \bibfield  {author} {\bibinfo {author} {\bibfnamefont {J.}~\bibnamefont
  {Zhu}}, \bibinfo {author} {\bibfnamefont {Y.}~\bibnamefont {Li}}, \bibinfo
  {author} {\bibfnamefont {X.}~\bibnamefont {Lin}}, \bibinfo {author}
  {\bibfnamefont {Y.}~\bibnamefont {Han}},\ and\ \bibinfo {author}
  {\bibfnamefont {K.}~\bibnamefont {Wu}},\ }\bibfield  {title} {\bibinfo
  {title} {Coherent phenomena and dynamics of lead halide perovskite
  nanocrystals for quantum information technologies},\ }\href
  {https://doi.org/10.1038/s41563-024-01922-z} {\bibfield  {journal} {\bibinfo
  {journal} {Nature Materials}\ }\textbf {\bibinfo {volume} {23}},\ \bibinfo
  {pages} {1027} (\bibinfo {year} {2024})}\BibitemShut {NoStop}%
\bibitem [{\citenamefont {Tamarat}\ \emph {et~al.}(2019)\citenamefont
  {Tamarat}, \citenamefont {Bodnarchuk}, \citenamefont {Trebbia}, \citenamefont
  {Erni}, \citenamefont {Kovalenko}, \citenamefont {Even},\ and\ \citenamefont
  {Lounis}}]{tamarat2019ground}%
  \BibitemOpen
  \bibfield  {author} {\bibinfo {author} {\bibfnamefont {P.}~\bibnamefont
  {Tamarat}}, \bibinfo {author} {\bibfnamefont {M.~I.}\ \bibnamefont
  {Bodnarchuk}}, \bibinfo {author} {\bibfnamefont {J.-B.}\ \bibnamefont
  {Trebbia}}, \bibinfo {author} {\bibfnamefont {R.}~\bibnamefont {Erni}},
  \bibinfo {author} {\bibfnamefont {M.~V.}\ \bibnamefont {Kovalenko}}, \bibinfo
  {author} {\bibfnamefont {J.}~\bibnamefont {Even}},\ and\ \bibinfo {author}
  {\bibfnamefont {B.}~\bibnamefont {Lounis}},\ }\bibfield  {title} {\bibinfo
  {title} {The ground exciton state of formamidinium lead bromide perovskite
  nanocrystals is a singlet dark state},\ }\href
  {https://doi.org/10.1038/s41563-019-0364-x} {\bibfield  {journal} {\bibinfo
  {journal} {Nature Materials}\ }\textbf {\bibinfo {volume} {18}},\ \bibinfo
  {pages} {717} (\bibinfo {year} {2019})}\BibitemShut {NoStop}%
\bibitem [{\citenamefont {Yu}(2016)}]{Yu2016}%
  \BibitemOpen
  \bibfield  {author} {\bibinfo {author} {\bibfnamefont {Z.~G.}\ \bibnamefont
  {Yu}},\ }\bibfield  {title} {\bibinfo {title} {{Effective-mass model and
  magneto-optical properties in hybrid perovskites}},\ }\href
  {https://doi.org/10.1038/srep28576} {\bibfield  {journal} {\bibinfo
  {journal} {Scientific Reports}\ }\textbf {\bibinfo {volume} {6}},\ \bibinfo
  {pages} {28576} (\bibinfo {year} {2016})}\BibitemShut {NoStop}%
\bibitem [{\citenamefont {Kirstein}\ \emph
  {et~al.}(2022{\natexlab{b}})\citenamefont {Kirstein}, \citenamefont
  {Yakovlev}, \citenamefont {Glazov}, \citenamefont {Evers}, \citenamefont
  {Zhukov}, \citenamefont {Belykh}, \citenamefont {Kopteva}, \citenamefont
  {Kudlacik}, \citenamefont {Nazarenko}, \citenamefont {Dirin}, \citenamefont
  {Kovalenko},\ and\ \citenamefont {Bayer}}]{Kirstein2022Adv}%
  \BibitemOpen
  \bibfield  {author} {\bibinfo {author} {\bibfnamefont {E.}~\bibnamefont
  {Kirstein}}, \bibinfo {author} {\bibfnamefont {D.~R.}\ \bibnamefont
  {Yakovlev}}, \bibinfo {author} {\bibfnamefont {M.~M.}\ \bibnamefont
  {Glazov}}, \bibinfo {author} {\bibfnamefont {E.}~\bibnamefont {Evers}},
  \bibinfo {author} {\bibfnamefont {E.~A.}\ \bibnamefont {Zhukov}}, \bibinfo
  {author} {\bibfnamefont {V.~V.}\ \bibnamefont {Belykh}}, \bibinfo {author}
  {\bibfnamefont {N.~E.}\ \bibnamefont {Kopteva}}, \bibinfo {author}
  {\bibfnamefont {D.}~\bibnamefont {Kudlacik}}, \bibinfo {author}
  {\bibfnamefont {O.}~\bibnamefont {Nazarenko}}, \bibinfo {author}
  {\bibfnamefont {D.~N.}\ \bibnamefont {Dirin}}, \bibinfo {author}
  {\bibfnamefont {M.~V.}\ \bibnamefont {Kovalenko}},\ and\ \bibinfo {author}
  {\bibfnamefont {M.}~\bibnamefont {Bayer}},\ }\bibfield  {title} {\bibinfo
  {title} {{Lead-dominated hyperfine interaction impacting the carrier spin
  dynamics in halide perovskites}},\ }\href
  {https://doi.org/10.1002/adma.202105263} {\bibfield  {journal} {\bibinfo
  {journal} {Advanced Materials}\ }\textbf {\bibinfo {volume} {34}},\ \bibinfo
  {pages} {2105263} (\bibinfo {year} {2022}{\natexlab{b}})}\BibitemShut
  {NoStop}%
\bibitem [{\citenamefont {Kotur}\ \emph {et~al.}(2026)\citenamefont {Kotur},
  \citenamefont {Bazhin}, \citenamefont {Kavokin}, \citenamefont {Kopteva},
  \citenamefont {Yakovlev}, \citenamefont {Kudlacik},\ and\ \citenamefont
  {Bayer}}]{kotur2026dynamic}%
  \BibitemOpen
  \bibfield  {author} {\bibinfo {author} {\bibfnamefont {M.}~\bibnamefont
  {Kotur}}, \bibinfo {author} {\bibfnamefont {P.~S.}\ \bibnamefont {Bazhin}},
  \bibinfo {author} {\bibfnamefont {K.~V.}\ \bibnamefont {Kavokin}}, \bibinfo
  {author} {\bibfnamefont {N.~E.}\ \bibnamefont {Kopteva}}, \bibinfo {author}
  {\bibfnamefont {D.~R.}\ \bibnamefont {Yakovlev}}, \bibinfo {author}
  {\bibfnamefont {D.}~\bibnamefont {Kudlacik}},\ and\ \bibinfo {author}
  {\bibfnamefont {M.}~\bibnamefont {Bayer}},\ }\bibfield  {title} {\bibinfo
  {title} {Dynamic polarization of nuclear spins by optically oriented
  electrons and holes in lead halide perovskite semiconductors},\ }\href
  {https://doi.org/10.1103/w11v-2v4g} {\bibfield  {journal} {\bibinfo
  {journal} {Physical Review B}\ }\textbf {\bibinfo {volume} {113}},\ \bibinfo
  {pages} {085204} (\bibinfo {year} {2026})}\BibitemShut {NoStop}%
\bibitem [{\citenamefont {Lin}\ \emph {et~al.}(2023)\citenamefont {Lin},
  \citenamefont {Han}, \citenamefont {Zhu},\ and\ \citenamefont
  {Wu}}]{lin2023room}%
  \BibitemOpen
  \bibfield  {author} {\bibinfo {author} {\bibfnamefont {X.}~\bibnamefont
  {Lin}}, \bibinfo {author} {\bibfnamefont {Y.}~\bibnamefont {Han}}, \bibinfo
  {author} {\bibfnamefont {J.}~\bibnamefont {Zhu}},\ and\ \bibinfo {author}
  {\bibfnamefont {K.}~\bibnamefont {Wu}},\ }\bibfield  {title} {\bibinfo
  {title} {Room-temperature coherent optical manipulation of hole spins in
  solution-grown perovskite quantum dots},\ }\href
  {https://doi.org/10.1038/s41565-022-01279-x} {\bibfield  {journal} {\bibinfo
  {journal} {Nature Nanotechnology}\ }\textbf {\bibinfo {volume} {18}},\
  \bibinfo {pages} {124} (\bibinfo {year} {2023})}\BibitemShut {NoStop}%
\bibitem [{\citenamefont {Meliakov}\ \emph {et~al.}(2023)\citenamefont
  {Meliakov}, \citenamefont {Zhukov}, \citenamefont {Kulebyakina},
  \citenamefont {Belykh},\ and\ \citenamefont {Yakovlev}}]{Meliakov2023}%
  \BibitemOpen
  \bibfield  {author} {\bibinfo {author} {\bibfnamefont {S.~R.}\ \bibnamefont
  {Meliakov}}, \bibinfo {author} {\bibfnamefont {E.~A.}\ \bibnamefont
  {Zhukov}}, \bibinfo {author} {\bibfnamefont {E.~V.}\ \bibnamefont
  {Kulebyakina}}, \bibinfo {author} {\bibfnamefont {V.~V.}\ \bibnamefont
  {Belykh}},\ and\ \bibinfo {author} {\bibfnamefont {D.~R.}\ \bibnamefont
  {Yakovlev}},\ }\bibfield  {title} {\bibinfo {title} {{Coherent Spin Dynamics
  of Electrons in CsPbBr$_3$ Perovskite Nanocrystals at Room Temperature}},\
  }\href {https://doi.org/10.3390/nano13172454} {\bibfield  {journal} {\bibinfo
   {journal} {Nanomaterials}\ }\textbf {\bibinfo {volume} {13}},\ \bibinfo
  {pages} {2454} (\bibinfo {year} {2023})}\BibitemShut {NoStop}%
\bibitem [{\citenamefont {Kopteva}\ \emph {et~al.}(2024)\citenamefont
  {Kopteva}, \citenamefont {Yakovlev}, \citenamefont {Yalcin}, \citenamefont
  {Akimov}, \citenamefont {Nestoklon}, \citenamefont {Glazov}, \citenamefont
  {Kotur}, \citenamefont {Kudlacik}, \citenamefont {Zhukov}, \citenamefont
  {Kirstein}, \citenamefont {Hordiichuk}, \citenamefont {Dirin}, \citenamefont
  {Kovalenko},\ and\ \citenamefont {Bayer}}]{Kopteva2024}%
  \BibitemOpen
  \bibfield  {author} {\bibinfo {author} {\bibfnamefont {N.~E.}\ \bibnamefont
  {Kopteva}}, \bibinfo {author} {\bibfnamefont {D.~R.}\ \bibnamefont
  {Yakovlev}}, \bibinfo {author} {\bibfnamefont {E.}~\bibnamefont {Yalcin}},
  \bibinfo {author} {\bibfnamefont {I.~A.}\ \bibnamefont {Akimov}}, \bibinfo
  {author} {\bibfnamefont {M.~O.}\ \bibnamefont {Nestoklon}}, \bibinfo {author}
  {\bibfnamefont {M.~M.}\ \bibnamefont {Glazov}}, \bibinfo {author}
  {\bibfnamefont {M.}~\bibnamefont {Kotur}}, \bibinfo {author} {\bibfnamefont
  {D.}~\bibnamefont {Kudlacik}}, \bibinfo {author} {\bibfnamefont {E.~A.}\
  \bibnamefont {Zhukov}}, \bibinfo {author} {\bibfnamefont {E.}~\bibnamefont
  {Kirstein}}, \bibinfo {author} {\bibfnamefont {O.}~\bibnamefont
  {Hordiichuk}}, \bibinfo {author} {\bibfnamefont {D.~N.}\ \bibnamefont
  {Dirin}}, \bibinfo {author} {\bibfnamefont {M.~V.}\ \bibnamefont
  {Kovalenko}},\ and\ \bibinfo {author} {\bibfnamefont {M.}~\bibnamefont
  {Bayer}},\ }\bibfield  {title} {\bibinfo {title} {{Highly-polarized emission
  provided by giant optical orientation of exciton spins in lead halide
  perovskite crystals}},\ }\href {https://doi.org/10.1002/advs.202403691}
  {\bibfield  {journal} {\bibinfo  {journal} {Advanced Science}\ }\textbf
  {\bibinfo {volume} {11}},\ \bibinfo {pages} {2403691} (\bibinfo {year}
  {2024})}\BibitemShut {NoStop}%
\bibitem [{\citenamefont {Kopteva}\ \emph {et~al.}(2025)\citenamefont
  {Kopteva}, \citenamefont {Yakovlev}, \citenamefont {Yalcin}, \citenamefont
  {Kalitukha}, \citenamefont {Akimov}, \citenamefont {Nestoklon}, \citenamefont
  {Turedi}, \citenamefont {Hordiichuk}, \citenamefont {Dirin}, \citenamefont
  {Kovalenko},\ and\ \citenamefont {Bayer}}]{Kopteva2025}%
  \BibitemOpen
  \bibfield  {author} {\bibinfo {author} {\bibfnamefont {N.~E.}\ \bibnamefont
  {Kopteva}}, \bibinfo {author} {\bibfnamefont {D.~R.}\ \bibnamefont
  {Yakovlev}}, \bibinfo {author} {\bibfnamefont {E.}~\bibnamefont {Yalcin}},
  \bibinfo {author} {\bibfnamefont {I.~V.}\ \bibnamefont {Kalitukha}}, \bibinfo
  {author} {\bibfnamefont {I.~A.}\ \bibnamefont {Akimov}}, \bibinfo {author}
  {\bibfnamefont {M.~O.}\ \bibnamefont {Nestoklon}}, \bibinfo {author}
  {\bibfnamefont {B.}~\bibnamefont {Turedi}}, \bibinfo {author} {\bibfnamefont
  {O.}~\bibnamefont {Hordiichuk}}, \bibinfo {author} {\bibfnamefont {D.~N.}\
  \bibnamefont {Dirin}}, \bibinfo {author} {\bibfnamefont {M.~V.}\ \bibnamefont
  {Kovalenko}},\ and\ \bibinfo {author} {\bibfnamefont {M.}~\bibnamefont
  {Bayer}},\ }\bibfield  {title} {\bibinfo {title} {{Effect of crystal symmetry
  of lead halide perovskites on the optical orientation of excitons}},\ }\href
  {https://doi.org/10.1002/advs.202416782} {\bibfield  {journal} {\bibinfo
  {journal} {Advanced Science}\ }\textbf {\bibinfo {volume} {12}},\ \bibinfo
  {pages} {2416782} (\bibinfo {year} {2025})}\BibitemShut {NoStop}%
\bibitem [{\citenamefont {Kudlacik}\ \emph {et~al.}(2024)\citenamefont
  {Kudlacik}, \citenamefont {Kopteva}, \citenamefont {Kotur}, \citenamefont
  {Yakovlev}, \citenamefont {Kavokin}, \citenamefont {Harkort}, \citenamefont
  {Karzel}, \citenamefont {Zhukov}, \citenamefont {Evers}, \citenamefont
  {Belykh},\ and\ \citenamefont {Bayer}}]{Kudlacik2024}%
  \BibitemOpen
  \bibfield  {author} {\bibinfo {author} {\bibfnamefont {D.}~\bibnamefont
  {Kudlacik}}, \bibinfo {author} {\bibfnamefont {N.~E.}\ \bibnamefont
  {Kopteva}}, \bibinfo {author} {\bibfnamefont {M.}~\bibnamefont {Kotur}},
  \bibinfo {author} {\bibfnamefont {D.~R.}\ \bibnamefont {Yakovlev}}, \bibinfo
  {author} {\bibfnamefont {K.~V.}\ \bibnamefont {Kavokin}}, \bibinfo {author}
  {\bibfnamefont {C.}~\bibnamefont {Harkort}}, \bibinfo {author} {\bibfnamefont
  {M.}~\bibnamefont {Karzel}}, \bibinfo {author} {\bibfnamefont {E.~A.}\
  \bibnamefont {Zhukov}}, \bibinfo {author} {\bibfnamefont {E.}~\bibnamefont
  {Evers}}, \bibinfo {author} {\bibfnamefont {V.~V.}\ \bibnamefont {Belykh}},\
  and\ \bibinfo {author} {\bibfnamefont {M.}~\bibnamefont {Bayer}},\ }\bibfield
   {title} {\bibinfo {title} {{Optical spin orientation of localized electrons
  and holes interacting with nuclei in a
  FA$_{0.9}$Cs$_{0.1}$PbI$_{2.8}$Br$_{0.2}$ perovskite crystal}},\ }\href
  {https://doi.org/10.1021/acsphotonics.4c00637} {\bibfield  {journal}
  {\bibinfo  {journal} {ACS Photonics}\ }\textbf {\bibinfo {volume} {11}},\
  \bibinfo {pages} {2757} (\bibinfo {year} {2024})}\BibitemShut {NoStop}%
\bibitem [{\citenamefont {Belykh}\ and\ \citenamefont
  {Melyakov}(2022)}]{Belykh2022PRB}%
  \BibitemOpen
  \bibfield  {author} {\bibinfo {author} {\bibfnamefont {V.~V.}\ \bibnamefont
  {Belykh}}\ and\ \bibinfo {author} {\bibfnamefont {S.~R.}\ \bibnamefont
  {Melyakov}},\ }\bibfield  {title} {\bibinfo {title} {{Selective measurement
  of the longitudinal electron spin relaxation time $T_1$ of Ce$^{3+}$ ions in
  a YAG lattice: Resonant spin inertia}},\ }\href
  {https://doi.org/10.1103/PhysRevB.105.205129} {\bibfield  {journal} {\bibinfo
   {journal} {Physical Review B}\ }\textbf {\bibinfo {volume} {105}},\ \bibinfo
  {pages} {205129} (\bibinfo {year} {2022})}\BibitemShut {NoStop}%
\bibitem [{\citenamefont {Belykh}\ \emph {et~al.}(2022)\citenamefont {Belykh},
  \citenamefont {Skorikov}, \citenamefont {Kulebyakina}, \citenamefont
  {Kolobkova}, \citenamefont {Kuznetsova}, \citenamefont {Glazov},\ and\
  \citenamefont {Yakovlev}}]{Belykh2022NL}%
  \BibitemOpen
  \bibfield  {author} {\bibinfo {author} {\bibfnamefont {V.~V.}\ \bibnamefont
  {Belykh}}, \bibinfo {author} {\bibfnamefont {M.~L.}\ \bibnamefont
  {Skorikov}}, \bibinfo {author} {\bibfnamefont {E.~V.}\ \bibnamefont
  {Kulebyakina}}, \bibinfo {author} {\bibfnamefont {E.~V.}\ \bibnamefont
  {Kolobkova}}, \bibinfo {author} {\bibfnamefont {M.~S.}\ \bibnamefont
  {Kuznetsova}}, \bibinfo {author} {\bibfnamefont {M.~M.}\ \bibnamefont
  {Glazov}},\ and\ \bibinfo {author} {\bibfnamefont {D.~R.}\ \bibnamefont
  {Yakovlev}},\ }\bibfield  {title} {\bibinfo {title} {{Submillisecond spin
  relaxation in CsPb(Cl,Br)$_3$ perovskite nanocrystals in a glass matrix}},\
  }\href {https://doi.org/10.1021/acs.nanolett.2c01673} {\bibfield  {journal}
  {\bibinfo  {journal} {Nano Letters}\ }\textbf {\bibinfo {volume} {22}},\
  \bibinfo {pages} {4583} (\bibinfo {year} {2022})}\BibitemShut {NoStop}%
\bibitem [{\citenamefont {Barak}\ \emph {et~al.}(2022)\citenamefont {Barak},
  \citenamefont {Meir}, \citenamefont {Dehnel}, \citenamefont {Horani},
  \citenamefont {Gamelin}, \citenamefont {Shapiro},\ and\ \citenamefont
  {Lifshitz}}]{barak2022uncovering}%
  \BibitemOpen
  \bibfield  {author} {\bibinfo {author} {\bibfnamefont {Y.}~\bibnamefont
  {Barak}}, \bibinfo {author} {\bibfnamefont {I.}~\bibnamefont {Meir}},
  \bibinfo {author} {\bibfnamefont {J.}~\bibnamefont {Dehnel}}, \bibinfo
  {author} {\bibfnamefont {F.}~\bibnamefont {Horani}}, \bibinfo {author}
  {\bibfnamefont {D.~R.}\ \bibnamefont {Gamelin}}, \bibinfo {author}
  {\bibfnamefont {A.}~\bibnamefont {Shapiro}},\ and\ \bibinfo {author}
  {\bibfnamefont {E.}~\bibnamefont {Lifshitz}},\ }\bibfield  {title} {\bibinfo
  {title} {Uncovering the influence of ni2+ doping in lead-halide perovskite
  nanocrystals using optically detected magnetic resonance spectroscopy},\
  }\href {https://doi.org/10.1021/acs.chemmater.1c03822} {\bibfield  {journal}
  {\bibinfo  {journal} {Chemistry of Materials}\ }\textbf {\bibinfo {volume}
  {34}},\ \bibinfo {pages} {1686} (\bibinfo {year} {2022})}\BibitemShut
  {NoStop}%
\bibitem [{\citenamefont {Bechtold}\ \emph {et~al.}(2015)\citenamefont
  {Bechtold}, \citenamefont {Rauch}, \citenamefont {Li}, \citenamefont
  {Simmet}, \citenamefont {Ardelt}, \citenamefont {Regler}, \citenamefont
  {M{\"{u}}ller}, \citenamefont {Sinitsyn},\ and\ \citenamefont
  {Finley}}]{Bechtold2015}%
  \BibitemOpen
  \bibfield  {author} {\bibinfo {author} {\bibfnamefont {A.}~\bibnamefont
  {Bechtold}}, \bibinfo {author} {\bibfnamefont {D.}~\bibnamefont {Rauch}},
  \bibinfo {author} {\bibfnamefont {F.}~\bibnamefont {Li}}, \bibinfo {author}
  {\bibfnamefont {T.}~\bibnamefont {Simmet}}, \bibinfo {author} {\bibfnamefont
  {P.-L.}\ \bibnamefont {Ardelt}}, \bibinfo {author} {\bibfnamefont
  {A.}~\bibnamefont {Regler}}, \bibinfo {author} {\bibfnamefont
  {K.}~\bibnamefont {M{\"{u}}ller}}, \bibinfo {author} {\bibfnamefont
  {N.}~\bibnamefont {Sinitsyn}},\ and\ \bibinfo {author} {\bibfnamefont
  {J.~J.}\ \bibnamefont {Finley}},\ }\bibfield  {title} {\bibinfo {title}
  {{Three-stage decoherence dynamics of an electron spin qubit in an optically
  active quantum dot}},\ }\href {https://doi.org/10.1038/nphys3470} {\bibfield
  {journal} {\bibinfo  {journal} {Nature Physics}\ }\textbf {\bibinfo {volume}
  {11}},\ \bibinfo {pages} {1005} (\bibinfo {year} {2015})}\BibitemShut
  {NoStop}%
\bibitem [{\citenamefont {Kirstein}\ \emph {et~al.}(2023)\citenamefont
  {Kirstein}, \citenamefont {Kopteva}, \citenamefont {Yakovlev}, \citenamefont
  {Zhukov}, \citenamefont {Kolobkova}, \citenamefont {Kuznetsova},
  \citenamefont {Belykh}, \citenamefont {Yugova}, \citenamefont {Glazov},
  \citenamefont {Bayer},\ and\ \citenamefont {Greilich}}]{Kirstein2023}%
  \BibitemOpen
  \bibfield  {author} {\bibinfo {author} {\bibfnamefont {E.}~\bibnamefont
  {Kirstein}}, \bibinfo {author} {\bibfnamefont {N.~E.}\ \bibnamefont
  {Kopteva}}, \bibinfo {author} {\bibfnamefont {D.~R.}\ \bibnamefont
  {Yakovlev}}, \bibinfo {author} {\bibfnamefont {E.~A.}\ \bibnamefont
  {Zhukov}}, \bibinfo {author} {\bibfnamefont {E.~V.}\ \bibnamefont
  {Kolobkova}}, \bibinfo {author} {\bibfnamefont {M.~S.}\ \bibnamefont
  {Kuznetsova}}, \bibinfo {author} {\bibfnamefont {V.~V.}\ \bibnamefont
  {Belykh}}, \bibinfo {author} {\bibfnamefont {I.~A.}\ \bibnamefont {Yugova}},
  \bibinfo {author} {\bibfnamefont {M.~M.}\ \bibnamefont {Glazov}}, \bibinfo
  {author} {\bibfnamefont {M.}~\bibnamefont {Bayer}},\ and\ \bibinfo {author}
  {\bibfnamefont {A.}~\bibnamefont {Greilich}},\ }\bibfield  {title} {\bibinfo
  {title} {{Mode locking of hole spin coherences in CsPb(Cl,Br)$_3$ perovskite
  nanocrystals}},\ }\href {https://doi.org/10.1038/s41467-023-36165-0}
  {\bibfield  {journal} {\bibinfo  {journal} {Nature Communications}\ }\textbf
  {\bibinfo {volume} {14}},\ \bibinfo {pages} {699} (\bibinfo {year}
  {2023})}\BibitemShut {NoStop}%
\bibitem [{\citenamefont {Meliakov}\ \emph {et~al.}(2024)\citenamefont
  {Meliakov}, \citenamefont {Zhukov}, \citenamefont {Belykh}, \citenamefont
  {Nestoklon}, \citenamefont {Kolobkova}, \citenamefont {Kuznetsova},
  \citenamefont {Bayer},\ and\ \citenamefont {Yakovlev}}]{Meliakov2024IT}%
  \BibitemOpen
  \bibfield  {author} {\bibinfo {author} {\bibfnamefont {S.~R.}\ \bibnamefont
  {Meliakov}}, \bibinfo {author} {\bibfnamefont {E.~A.}\ \bibnamefont
  {Zhukov}}, \bibinfo {author} {\bibfnamefont {V.~V.}\ \bibnamefont {Belykh}},
  \bibinfo {author} {\bibfnamefont {M.~O.}\ \bibnamefont {Nestoklon}}, \bibinfo
  {author} {\bibfnamefont {E.~V.}\ \bibnamefont {Kolobkova}}, \bibinfo {author}
  {\bibfnamefont {M.~S.}\ \bibnamefont {Kuznetsova}}, \bibinfo {author}
  {\bibfnamefont {M.}~\bibnamefont {Bayer}},\ and\ \bibinfo {author}
  {\bibfnamefont {D.~R.}\ \bibnamefont {Yakovlev}},\ }\bibfield  {title}
  {\bibinfo {title} {{Temperature dependence of the electron and hole
  Land{\'{e}} $g$-factors in CsPbI$_3$ nanocrystals embedded in a glass
  matrix}},\ }\href {https://doi.org/10.1039/D4NR03132F} {\bibfield  {journal}
  {\bibinfo  {journal} {Nanoscale}\ }\textbf {\bibinfo {volume} {16}},\
  \bibinfo {pages} {21496} (\bibinfo {year} {2024})}\BibitemShut {NoStop}%
\bibitem [{\citenamefont {Heisterkamp}\ \emph {et~al.}(2015)\citenamefont
  {Heisterkamp}, \citenamefont {Zhukov}, \citenamefont {Greilich},
  \citenamefont {Yakovlev}, \citenamefont {Korenev}, \citenamefont {Pawlis},\
  and\ \citenamefont {Bayer}}]{Heisterkamp2015}%
  \BibitemOpen
  \bibfield  {author} {\bibinfo {author} {\bibfnamefont {F.}~\bibnamefont
  {Heisterkamp}}, \bibinfo {author} {\bibfnamefont {E.~A.}\ \bibnamefont
  {Zhukov}}, \bibinfo {author} {\bibfnamefont {A.}~\bibnamefont {Greilich}},
  \bibinfo {author} {\bibfnamefont {D.~R.}\ \bibnamefont {Yakovlev}}, \bibinfo
  {author} {\bibfnamefont {V.~L.}\ \bibnamefont {Korenev}}, \bibinfo {author}
  {\bibfnamefont {A.}~\bibnamefont {Pawlis}},\ and\ \bibinfo {author}
  {\bibfnamefont {M.}~\bibnamefont {Bayer}},\ }\bibfield  {title} {\bibinfo
  {title} {{Longitudinal and transverse spin dynamics of donor-bound electrons
  in fluorine-doped ZnSe: Spin inertia versus Hanle effect}},\ }\href
  {https://doi.org/10.1103/PhysRevB.91.235432} {\bibfield  {journal} {\bibinfo
  {journal} {Physical Review B}\ }\textbf {\bibinfo {volume} {91}},\ \bibinfo
  {pages} {235432} (\bibinfo {year} {2015})}\BibitemShut {NoStop}%
\bibitem [{\citenamefont {Belykh}\ \emph {et~al.}(2016)\citenamefont {Belykh},
  \citenamefont {Yakovlev}, \citenamefont {Schindler}, \citenamefont {Zhukov},
  \citenamefont {Semina}, \citenamefont {Yacob}, \citenamefont {Reithmaier},
  \citenamefont {Benyoucef},\ and\ \citenamefont {Bayer}}]{belykh2016large}%
  \BibitemOpen
  \bibfield  {author} {\bibinfo {author} {\bibfnamefont {V.~V.}\ \bibnamefont
  {Belykh}}, \bibinfo {author} {\bibfnamefont {D.~R.}\ \bibnamefont
  {Yakovlev}}, \bibinfo {author} {\bibfnamefont {J.~J.}\ \bibnamefont
  {Schindler}}, \bibinfo {author} {\bibfnamefont {E.~A.}\ \bibnamefont
  {Zhukov}}, \bibinfo {author} {\bibfnamefont {M.~A.}\ \bibnamefont {Semina}},
  \bibinfo {author} {\bibfnamefont {M.}~\bibnamefont {Yacob}}, \bibinfo
  {author} {\bibfnamefont {J.~P.}\ \bibnamefont {Reithmaier}}, \bibinfo
  {author} {\bibfnamefont {M.}~\bibnamefont {Benyoucef}},\ and\ \bibinfo
  {author} {\bibfnamefont {M.}~\bibnamefont {Bayer}},\ }\bibfield  {title}
  {\bibinfo {title} {Large anisotropy of electron and hole g factors in
  infrared-emitting inas/inalgaas self-assembled quantum dots},\ }\href
  {https://doi.org/10.1103/PhysRevB.93.125302} {\bibfield  {journal} {\bibinfo
  {journal} {Physical Review B}\ }\textbf {\bibinfo {volume} {93}},\ \bibinfo
  {pages} {125302} (\bibinfo {year} {2016})}\BibitemShut {NoStop}%
\bibitem [{\citenamefont {Merkulov}\ \emph {et~al.}(2002)\citenamefont
  {Merkulov}, \citenamefont {Efros},\ and\ \citenamefont
  {Rosen}}]{merkulov2002electron}%
  \BibitemOpen
  \bibfield  {author} {\bibinfo {author} {\bibfnamefont {I.}~\bibnamefont
  {Merkulov}}, \bibinfo {author} {\bibfnamefont {A.~L.}\ \bibnamefont
  {Efros}},\ and\ \bibinfo {author} {\bibfnamefont {M.}~\bibnamefont {Rosen}},\
  }\bibfield  {title} {\bibinfo {title} {Electron spin relaxation by nuclei in
  semiconductor quantum dots},\ }\href
  {https://doi.org/10.1103/PhysRevB.65.205309} {\bibfield  {journal} {\bibinfo
  {journal} {Physical review B}\ }\textbf {\bibinfo {volume} {65}},\ \bibinfo
  {pages} {205309} (\bibinfo {year} {2002})}\BibitemShut {NoStop}%
\bibitem [{\citenamefont {Glazov}(2018)}]{glazovElectronNuclearSpin2018}%
  \BibitemOpen
  \bibfield  {author} {\bibinfo {author} {\bibfnamefont {M.~M.}\ \bibnamefont
  {Glazov}},\ }\href@noop {} {\emph {\bibinfo {title} {Electron \& {{Nuclear
  Spin Dynamics}} in {{semiconductor Nanostructures}}}}},\ Series on
  {{Semiconductor Science}} and {{Technology}}\ (\bibinfo  {publisher} {OUP
  Oxford},\ \bibinfo {year} {2018})\BibitemShut {NoStop}%
\bibitem [{\citenamefont {Meliakov}\ \emph {et~al.}(2026)\citenamefont
  {Meliakov}, \citenamefont {Zhukov}, \citenamefont {Belykh}, \citenamefont
  {Kavokin}, \citenamefont {Nestoklon}, \citenamefont {Kulebyakina},
  \citenamefont {Skorikov}, \citenamefont {Kolobkova}, \citenamefont
  {Kuznetsova}, \citenamefont {Bayer},\ and\ \citenamefont
  {Yakovlev}}]{Meliakov2026Nucl}%
  \BibitemOpen
  \bibfield  {author} {\bibinfo {author} {\bibfnamefont {S.~R.}\ \bibnamefont
  {Meliakov}}, \bibinfo {author} {\bibfnamefont {E.~A.}\ \bibnamefont
  {Zhukov}}, \bibinfo {author} {\bibfnamefont {V.~V.}\ \bibnamefont {Belykh}},
  \bibinfo {author} {\bibfnamefont {K.~V.}\ \bibnamefont {Kavokin}}, \bibinfo
  {author} {\bibfnamefont {M.~O.}\ \bibnamefont {Nestoklon}}, \bibinfo {author}
  {\bibfnamefont {E.~V.}\ \bibnamefont {Kulebyakina}}, \bibinfo {author}
  {\bibfnamefont {M.~L.}\ \bibnamefont {Skorikov}}, \bibinfo {author}
  {\bibfnamefont {E.~V.}\ \bibnamefont {Kolobkova}}, \bibinfo {author}
  {\bibfnamefont {M.~S.}\ \bibnamefont {Kuznetsova}}, \bibinfo {author}
  {\bibfnamefont {M.}~\bibnamefont {Bayer}},\ and\ \bibinfo {author}
  {\bibfnamefont {D.~R.}\ \bibnamefont {Yakovlev}},\ }\bibfield  {title}
  {\bibinfo {title} {{Hyperfine interaction of electrons confined in CsPbI$_3$
  nanocrystals with nuclear spin fluctuations}},\ }\href
  {https://doi.org/10.1103/s33c-m6hz} {\bibfield  {journal} {\bibinfo
  {journal} {Physical Review B}\ }\textbf {\bibinfo {volume} {113}},\ \bibinfo
  {pages} {035304} (\bibinfo {year} {2026})}\BibitemShut {NoStop}%
\bibitem [{\citenamefont {Harkort}\ \emph {et~al.}(2025)\citenamefont
  {Harkort}, \citenamefont {Kalitukha}, \citenamefont {Kopteva}, \citenamefont
  {Nestoklon}, \citenamefont {Goupalov}, \citenamefont {Saviot}, \citenamefont
  {Kudlacik}, \citenamefont {Yakovlev}, \citenamefont {Kolobkova},
  \citenamefont {Kuznetsova},\ and\ \citenamefont {Bayer}}]{Harkort2025}%
  \BibitemOpen
  \bibfield  {author} {\bibinfo {author} {\bibfnamefont {C.}~\bibnamefont
  {Harkort}}, \bibinfo {author} {\bibfnamefont {I.~V.}\ \bibnamefont
  {Kalitukha}}, \bibinfo {author} {\bibfnamefont {N.~E.}\ \bibnamefont
  {Kopteva}}, \bibinfo {author} {\bibfnamefont {M.~O.}\ \bibnamefont
  {Nestoklon}}, \bibinfo {author} {\bibfnamefont {S.~V.}\ \bibnamefont
  {Goupalov}}, \bibinfo {author} {\bibfnamefont {L.}~\bibnamefont {Saviot}},
  \bibinfo {author} {\bibfnamefont {D.}~\bibnamefont {Kudlacik}}, \bibinfo
  {author} {\bibfnamefont {D.~R.}\ \bibnamefont {Yakovlev}}, \bibinfo {author}
  {\bibfnamefont {E.~V.}\ \bibnamefont {Kolobkova}}, \bibinfo {author}
  {\bibfnamefont {M.~S.}\ \bibnamefont {Kuznetsova}},\ and\ \bibinfo {author}
  {\bibfnamefont {M.}~\bibnamefont {Bayer}},\ }\bibfield  {title} {\bibinfo
  {title} {{Confined acoustic phonons in CsPbI$_3$ nanocrystals explored by
  resonant raman scattering on excitons}},\ }\href
  {https://doi.org/10.1021/acs.nanolett.5c03342} {\bibfield  {journal}
  {\bibinfo  {journal} {Nano Letters}\ }\textbf {\bibinfo {volume} {25}},\
  \bibinfo {pages} {12754} (\bibinfo {year} {2025})}\BibitemShut {NoStop}%
\bibitem [{\citenamefont {Kulebyakina}\ \emph {et~al.}(2024)\citenamefont
  {Kulebyakina}, \citenamefont {Skorikov}, \citenamefont {Kolobkova},
  \citenamefont {Kuznetsova}, \citenamefont {Bataev}, \citenamefont
  {Yakovlev},\ and\ \citenamefont {Belykh}}]{kulebyakina2024temperature}%
  \BibitemOpen
  \bibfield  {author} {\bibinfo {author} {\bibfnamefont {E.~V.}\ \bibnamefont
  {Kulebyakina}}, \bibinfo {author} {\bibfnamefont {M.~L.}\ \bibnamefont
  {Skorikov}}, \bibinfo {author} {\bibfnamefont {E.~V.}\ \bibnamefont
  {Kolobkova}}, \bibinfo {author} {\bibfnamefont {M.~S.}\ \bibnamefont
  {Kuznetsova}}, \bibinfo {author} {\bibfnamefont {M.~N.}\ \bibnamefont
  {Bataev}}, \bibinfo {author} {\bibfnamefont {D.~R.}\ \bibnamefont
  {Yakovlev}},\ and\ \bibinfo {author} {\bibfnamefont {V.~V.}\ \bibnamefont
  {Belykh}},\ }\bibfield  {title} {\bibinfo {title} {Temperature-dependent
  photoluminescence dynamics of cspbbr$_3$ and cspb(cl,br)$_3$ perovskite
  nanocrystals in a glass matrix},\ }\href
  {https://doi.org/10.1103/PhysRevB.109.235301} {\bibfield  {journal} {\bibinfo
   {journal} {Physical Review B}\ }\textbf {\bibinfo {volume} {109}},\ \bibinfo
  {pages} {235301} (\bibinfo {year} {2024})}\BibitemShut {NoStop}%
\bibitem [{\citenamefont {Greilich}\ \emph {et~al.}(2006)\citenamefont
  {Greilich}, \citenamefont {Oulton}, \citenamefont {Zhukov}, \citenamefont
  {Yugova}, \citenamefont {Yakovlev}, \citenamefont {Bayer}, \citenamefont
  {Shabaev}, \citenamefont {Efros}, \citenamefont {Merkulov}, \citenamefont
  {Stavarache}, \citenamefont {Reuter},\ and\ \citenamefont
  {Wieck}}]{Greilich2006}%
  \BibitemOpen
  \bibfield  {author} {\bibinfo {author} {\bibfnamefont {A.}~\bibnamefont
  {Greilich}}, \bibinfo {author} {\bibfnamefont {R.}~\bibnamefont {Oulton}},
  \bibinfo {author} {\bibfnamefont {E.~A.}\ \bibnamefont {Zhukov}}, \bibinfo
  {author} {\bibfnamefont {I.~A.}\ \bibnamefont {Yugova}}, \bibinfo {author}
  {\bibfnamefont {D.~R.}\ \bibnamefont {Yakovlev}}, \bibinfo {author}
  {\bibfnamefont {M.}~\bibnamefont {Bayer}}, \bibinfo {author} {\bibfnamefont
  {A.}~\bibnamefont {Shabaev}}, \bibinfo {author} {\bibfnamefont {A.~L.}\
  \bibnamefont {Efros}}, \bibinfo {author} {\bibfnamefont {I.~A.}\ \bibnamefont
  {Merkulov}}, \bibinfo {author} {\bibfnamefont {V.}~\bibnamefont
  {Stavarache}}, \bibinfo {author} {\bibfnamefont {D.}~\bibnamefont {Reuter}},\
  and\ \bibinfo {author} {\bibfnamefont {A.}~\bibnamefont {Wieck}},\ }\bibfield
   {title} {\bibinfo {title} {{Optical control of spin coherence in singly
  charged (In,Ga)As/GaAs quantum dots}},\ }\href
  {https://doi.org/10.1103/PhysRevLett.96.227401} {\bibfield  {journal}
  {\bibinfo  {journal} {Physical Review Letters}\ }\textbf {\bibinfo {volume}
  {96}},\ \bibinfo {pages} {227401} (\bibinfo {year} {2006})}\BibitemShut
  {NoStop}%
\bibitem [{\citenamefont {Smirnov}\ \emph {et~al.}(2018)\citenamefont
  {Smirnov}, \citenamefont {Zhukov}, \citenamefont {Kirstein}, \citenamefont
  {Yakovlev}, \citenamefont {Reuter}, \citenamefont {Wieck}, \citenamefont
  {Bayer}, \citenamefont {Greilich},\ and\ \citenamefont
  {Glazov}}]{Smirnov2018}%
  \BibitemOpen
  \bibfield  {author} {\bibinfo {author} {\bibfnamefont {D.~S.}\ \bibnamefont
  {Smirnov}}, \bibinfo {author} {\bibfnamefont {E.~A.}\ \bibnamefont {Zhukov}},
  \bibinfo {author} {\bibfnamefont {E.}~\bibnamefont {Kirstein}}, \bibinfo
  {author} {\bibfnamefont {D.~R.}\ \bibnamefont {Yakovlev}}, \bibinfo {author}
  {\bibfnamefont {D.}~\bibnamefont {Reuter}}, \bibinfo {author} {\bibfnamefont
  {A.~D.}\ \bibnamefont {Wieck}}, \bibinfo {author} {\bibfnamefont
  {M.}~\bibnamefont {Bayer}}, \bibinfo {author} {\bibfnamefont
  {A.}~\bibnamefont {Greilich}},\ and\ \bibinfo {author} {\bibfnamefont
  {M.~M.}\ \bibnamefont {Glazov}},\ }\bibfield  {title} {\bibinfo {title}
  {{Theory of spin inertia in singly charged quantum dots}},\ }\href
  {https://doi.org/10.1103/PhysRevB.98.125306} {\bibfield  {journal} {\bibinfo
  {journal} {Physical Review B}\ }\textbf {\bibinfo {volume} {98}},\ \bibinfo
  {pages} {125306} (\bibinfo {year} {2018})}\BibitemShut {NoStop}%
\bibitem [{\citenamefont {Trifonov}\ \emph {et~al.}(2026)\citenamefont
  {Trifonov}, \citenamefont {Nestoklon}, \citenamefont {Hollberg},
  \citenamefont {Grisard}, \citenamefont {Kudlacik}, \citenamefont {Kolobkova},
  \citenamefont {Kuznetsova}, \citenamefont {Goupalov}, \citenamefont
  {Kaspari}, \citenamefont {Reiter}, \citenamefont {Yakovlev}, \citenamefont
  {Bayer},\ and\ \citenamefont {Akimov}}]{Trifonov2026}%
  \BibitemOpen
  \bibfield  {author} {\bibinfo {author} {\bibfnamefont {A.~V.}\ \bibnamefont
  {Trifonov}}, \bibinfo {author} {\bibfnamefont {M.~O.}\ \bibnamefont
  {Nestoklon}}, \bibinfo {author} {\bibfnamefont {M.~A.}\ \bibnamefont
  {Hollberg}}, \bibinfo {author} {\bibfnamefont {S.}~\bibnamefont {Grisard}},
  \bibinfo {author} {\bibfnamefont {D.}~\bibnamefont {Kudlacik}}, \bibinfo
  {author} {\bibfnamefont {E.~V.}\ \bibnamefont {Kolobkova}}, \bibinfo {author}
  {\bibfnamefont {M.~S.}\ \bibnamefont {Kuznetsova}}, \bibinfo {author}
  {\bibfnamefont {S.~V.}\ \bibnamefont {Goupalov}}, \bibinfo {author}
  {\bibfnamefont {J.~M.}\ \bibnamefont {Kaspari}}, \bibinfo {author}
  {\bibfnamefont {D.~E.}\ \bibnamefont {Reiter}}, \bibinfo {author}
  {\bibfnamefont {D.~R.}\ \bibnamefont {Yakovlev}}, \bibinfo {author}
  {\bibfnamefont {M.}~\bibnamefont {Bayer}},\ and\ \bibinfo {author}
  {\bibfnamefont {I.~A.}\ \bibnamefont {Akimov}},\ }\bibfield  {title}
  {\bibinfo {title} {{Quantum beats of exciton-polarons in CsPbI3 perovskite
  nanocrystals}},\ }\href {https://doi.org/10.1038/s41467-026-73506-1}
  {\bibfield  {journal} {\bibinfo  {journal} {Nature Communications}\ }\textbf
  {\bibinfo {volume} {17}},\ \bibinfo {pages} {4685} (\bibinfo {year}
  {2026})},\ \Eprint {https://arxiv.org/abs/2510.14695} {arXiv:2510.14695}
  \BibitemShut {NoStop}%
\bibitem [{\citenamefont {Pavlov}\ and\ \citenamefont
  {Firsov}(1966)}]{pavlovSpinFlipInteraction1966}%
  \BibitemOpen
  \bibfield  {author} {\bibinfo {author} {\bibfnamefont {S.~T.}\ \bibnamefont
  {Pavlov}}\ and\ \bibinfo {author} {\bibfnamefont {Y.~A.}\ \bibnamefont
  {Firsov}},\ }\bibfield  {title} {\bibinfo {title} {Spin flip interaction of
  electrons with optical phonons in semiconductors},\ }\href@noop {} {\bibfield
   {journal} {\bibinfo  {journal} {Sov. Phys. Solid State}\ }\textbf {\bibinfo
  {volume} {7}},\ \bibinfo {pages} {2131} (\bibinfo {year} {1966})}\BibitemShut
  {NoStop}%
\bibitem [{\citenamefont {Pavlov}\ and\ \citenamefont
  {Firsov}(1967)}]{pavlovSpinphononInteractionElectrons1967}%
  \BibitemOpen
  \bibfield  {author} {\bibinfo {author} {\bibfnamefont {S.~T.}\ \bibnamefont
  {Pavlov}}\ and\ \bibinfo {author} {\bibfnamefont {Y.~A.}\ \bibnamefont
  {Firsov}},\ }\bibfield  {title} {\bibinfo {title} {On spin-phonon interaction
  of electrons and oscillations of longitudinal magnetoresistance in
  semiconductors},\ }\href@noop {} {\bibfield  {journal} {\bibinfo  {journal}
  {Sov. Phys. Solid State}\ }\textbf {\bibinfo {volume} {9}},\ \bibinfo {pages}
  {1394} (\bibinfo {year} {1967})}\BibitemShut {NoStop}%
\bibitem [{\citenamefont {Englman}\ and\ \citenamefont
  {Ruppin}(1968)}]{englmanOpticalLatticeVibrations1968a}%
  \BibitemOpen
  \bibfield  {author} {\bibinfo {author} {\bibfnamefont {R.}~\bibnamefont
  {Englman}}\ and\ \bibinfo {author} {\bibfnamefont {R.}~\bibnamefont
  {Ruppin}},\ }\bibfield  {title} {\bibinfo {title} {Optical lattice vibrations
  in finite ionic crystals: {{I}}},\ }\href
  {https://doi.org/10.1088/0022-3719/1/3/309} {\bibfield  {journal} {\bibinfo
  {journal} {Journal of Physics C: Solid State Physics}\ }\textbf {\bibinfo
  {volume} {1}},\ \bibinfo {pages} {614} (\bibinfo {year} {1968})}\BibitemShut
  {NoStop}%
\bibitem [{\citenamefont {Ruppin}\ and\ \citenamefont
  {Englman}(1970)}]{ruppinOpticalPhononsSmall1970}%
  \BibitemOpen
  \bibfield  {author} {\bibinfo {author} {\bibfnamefont {R.}~\bibnamefont
  {Ruppin}}\ and\ \bibinfo {author} {\bibfnamefont {R.}~\bibnamefont
  {Englman}},\ }\bibfield  {title} {\bibinfo {title} {Optical phonons of small
  crystals},\ }\href {https://doi.org/10.1088/0034-4885/33/1/304} {\bibfield
  {journal} {\bibinfo  {journal} {Reports on Progress in Physics}\ }\textbf
  {\bibinfo {volume} {33}},\ \bibinfo {pages} {149} (\bibinfo {year}
  {1970})}\BibitemShut {NoStop}%
\bibitem [{\citenamefont {Efros}(1993)}]{efrosElectronHolePairPhonon1993}%
  \BibitemOpen
  \bibfield  {author} {\bibinfo {author} {\bibfnamefont {{\relax Al}.~L.}\
  \bibnamefont {Efros}},\ }\bibfield  {title} {\bibinfo {title}
  {Electron-{{hole pair}} --- {{phonon interaction}} in {{semiconductor
  microcrystals}}},\ }in\ \href {https://doi.org/10.1007/978-94-011-1683-1_29}
  {\emph {\bibinfo {booktitle} {Phonons in {{Semiconductor
  Nanostructures}}}}},\ \bibinfo {editor} {edited by\ \bibinfo {editor}
  {\bibfnamefont {J.-P.}\ \bibnamefont {Leburton}}, \bibinfo {editor}
  {\bibfnamefont {J.}~\bibnamefont {Pascual}},\ and\ \bibinfo {editor}
  {\bibfnamefont {C.~S.}\ \bibnamefont {Torres}}}\ (\bibinfo  {publisher}
  {Springer Netherlands},\ \bibinfo {address} {Dordrecht},\ \bibinfo {year}
  {1993})\ pp.\ \bibinfo {pages} {299--308}\BibitemShut {NoStop}%
\bibitem [{\citenamefont {Klein}\ \emph {et~al.}(1990)\citenamefont {Klein},
  \citenamefont {Hache}, \citenamefont {Ricard},\ and\ \citenamefont
  {Flytzanis}}]{kleinSizeDependenceElectronphonon1990}%
  \BibitemOpen
  \bibfield  {author} {\bibinfo {author} {\bibfnamefont {M.~C.}\ \bibnamefont
  {Klein}}, \bibinfo {author} {\bibfnamefont {F.}~\bibnamefont {Hache}},
  \bibinfo {author} {\bibfnamefont {D.}~\bibnamefont {Ricard}},\ and\ \bibinfo
  {author} {\bibfnamefont {C.}~\bibnamefont {Flytzanis}},\ }\bibfield  {title}
  {\bibinfo {title} {Size dependence of electron-phonon coupling in
  semiconductor nanospheres: {{the}} case of {{CdSe}}},\ }\href
  {https://doi.org/10.1103/PhysRevB.42.11123} {\bibfield  {journal} {\bibinfo
  {journal} {Phys. Rev. B}\ }\textbf {\bibinfo {volume} {42}},\ \bibinfo
  {pages} {11123} (\bibinfo {year} {1990})}\BibitemShut {NoStop}%
\bibitem [{\citenamefont {Prokofiev}\ \emph {et~al.}(2014)\citenamefont
  {Prokofiev}, \citenamefont {Poddubny},\ and\ \citenamefont
  {Yassievich}}]{prokofiev2014phonon}%
  \BibitemOpen
  \bibfield  {author} {\bibinfo {author} {\bibfnamefont {A.~A.}\ \bibnamefont
  {Prokofiev}}, \bibinfo {author} {\bibfnamefont {A.~N.}\ \bibnamefont
  {Poddubny}},\ and\ \bibinfo {author} {\bibfnamefont {I.~N.}\ \bibnamefont
  {Yassievich}},\ }\bibfield  {title} {\bibinfo {title} {Phonon decay in
  silicon nanocrystals: Fast phonon recycling},\ }\href
  {https://doi.org/10.1103/PhysRevB.89.125409} {\bibfield  {journal} {\bibinfo
  {journal} {Physical Review B}\ }\textbf {\bibinfo {volume} {89}},\ \bibinfo
  {pages} {125409} (\bibinfo {year} {2014})}\BibitemShut {NoStop}%
\bibitem [{\citenamefont {Winkler}(2003)}]{winklerSpinOrbitCoupling2003}%
  \BibitemOpen
  \bibfield  {author} {\bibinfo {author} {\bibfnamefont {R.}~\bibnamefont
  {Winkler}},\ }\href@noop {} {\emph {\bibinfo {title} {Spin--{{Orbit Coupling
  Effects}} in {{Two-Dimensional Electron}} and {{Hole Systems}}}}}\ (\bibinfo
  {publisher} {Springer},\ \bibinfo {year} {2003})\BibitemShut {NoStop}%
\bibitem [{\citenamefont {{Andrada e Silva}}\ \emph {et~al.}(1994)\citenamefont
  {{Andrada e Silva}}, \citenamefont {Rocca},\ and\ \citenamefont
  {Bassani}}]{andradaesilvaSpinsplitSubbandsMagnetooscillations1994}%
  \BibitemOpen
  \bibfield  {author} {\bibinfo {author} {\bibfnamefont {E.~A.}\ \bibnamefont
  {{Andrada e Silva}}}, \bibinfo {author} {\bibfnamefont {G.~C.~L.}\
  \bibnamefont {Rocca}},\ and\ \bibinfo {author} {\bibfnamefont
  {F.}~\bibnamefont {Bassani}},\ }\bibfield  {title} {\bibinfo {title}
  {Spin-split subbands and magneto-oscillations in {{III-V}} asymetric
  heterostructures},\ }\href {https://doi.org/10.1103/PhysRevB.50.8523}
  {\bibfield  {journal} {\bibinfo  {journal} {Phys. Rev. B}\ }\textbf {\bibinfo
  {volume} {50}},\ \bibinfo {pages} {8523} (\bibinfo {year}
  {1994})}\BibitemShut {NoStop}%
\bibitem [{\citenamefont {Landau}\ and\ \citenamefont
  {Lifshitz}(1977)}]{landauQuantumMechanicsNonRelativistic1977}%
  \BibitemOpen
  \bibfield  {author} {\bibinfo {author} {\bibfnamefont {L.~D.}\ \bibnamefont
  {Landau}}\ and\ \bibinfo {author} {\bibfnamefont {E.~M.}\ \bibnamefont
  {Lifshitz}},\ }\href@noop {} {\emph {\bibinfo {title} {Quantum {{Mechanics}}:
  {{Non-Relativistic Theory}}}}}\ (\bibinfo  {publisher}
  {Butterworth-Heinemann, Oxford},\ \bibinfo {year} {1977})\BibitemShut
  {NoStop}%
\end{thebibliography}
\end{document}